\pdfoutput=1
\documentclass{JINST}
\let\ifpdf\relax
\usepackage{epsfig}

\title {Annealing effects on operation of thin Low Gain Avalanche Detectors \thanks{Work performed in the framework of the CERN-RD50 collaboration.}} 
\author{G.\,Kramberger$^{a}$\thanks{Corresponding author; E-mail:Gregor.Kramberger@ijs.si, Tel: (+386) 1 477 3512}, V.\,Cindro$^{a}$, A. Howard$^{a}$, \v Z. Kljun$^{a}$, I.\,Mandi\' c$^{a}$, M.\,Miku\v z$^{a}$$^{b}$  \\
\llap{$^{a}$}  Jo\v zef Stefan Institute, Jamova 39, SI-1000 Ljubljana, Slovenia \\
\llap{$^{b}$}  University of Ljubljana, Faculty of Mathematics and Physics, Jadranska 19, SI-1000 Ljubljana, Slovenia \\
}

\abstract{ Several thin Low Gain Avalanche Detectors from Hamamatsu Photonics were irradiated with neutrons
to different equivalent fluences up to $\Phi_{eq}=3\cdot10^{15}$ cm$^{-2}$. After the irradiation they were
annealed at 60$^\circ$C in steps to times $>20000$ minutes. Their properties, mainly full depletion voltage,
gain layer depletion voltage, generation and leakage current, as well as their performance in terms of
collected charge and time resolution, were determined between the steps.

It was found that the effect of annealing on timing resolution and collected charge is not very large and mainly
occurs within the first few tens of minutes. It is a consequence of active initial acceptor concentration decrease in the
 gain layer with time, where changes of around 10\% were observed. For any relevant annealing times for detector operation the changes of
effective doping concentration in the bulk negligibly influences the performance of the device, due to their small thickness and required
high bias voltage operation. At very long annealing times the
increase of the effective doping  concentration in the bulk leads to a significant increase of the electric field in the gain layer and,
by that, to the increase of gain at given voltage. The leakage current decreases in accordance with generation current annealing. 
}

\begin{document}

\section{Introduction}

Low gain avalanche detectors (LGAD) are going to be used as timing detectors at the upgraded
Large Hadron Collider (HL-LHC) at CERN after 2027, due to their excellent timing resolution of 20-30 ps \cite{UFSD1}. 
LGADs exploit a n$^{++}$-p$^{+}$-p-p$^{++}$ structure to achieve high enough electric fields near the junction contact 
for impact ionization \cite{LGAD-Main}. The gain depends on the p$^+$ layer's doping level and profile shape. The
doping levels of $\sim 10^{16}$ cm$^{-3}$ and implant widths of 1-2 $\mu$m at depths of up to 3 $\mu$m are used.
Usually gain factors of several tens were achieved in most LGADs produced so far. The main obstacle for their use
is the decrease of gain with radiation, which deactivates initial acceptors in the gain layer. The
most efficient ways to mitigate the gain decrease so far are using deeper profiles and Carbon
co-implantation \cite{Carbon}, but the radiation damage dictates three sensor replacements during the lifetime
of ATLAS High Granularity Timing Detector (ATLAS-HGTD)\cite{ATLAS}.
The studies performed so far were almost exclusively done after standard annealing time of 80 min at 60$^\circ$C, which is
roughly the time required to complete short term annealing of radiation induced deep acceptors and corresponds to 
the time the detectors will be kept at room temperature during yearly maintenance period.

In order to plan a running scenario and foresee operation in case of unplanned events it is important to quantify the effects of
annealing to LGAD operation. The annealing of detector bulk is expected to be similar to the one of standard 
n-p silicon detectors while little is known of gain layer annealing. It is important to understand how gain layer doping
changes with time and to establish the underlying reason for that. In addition, the effects the bulk
doping has on the electric field in the gain layer, and the possible effects of enhanced free hole concentration originating in the
gain layer to the electric field, have to be addressed. Equally important is the generation current annealing which to a large
extent steers the leakage current. 

It is therefore the intention of this paper to investigate all the effects of annealing on LGAD operation.

\section {Annealing effects}

The effective doping concentration (negative) in the gain layer $N_{gl}$ is given as the sum of deep ($N_{deep}$)
and initial dopants ($N_B$) and can be expressed as \cite{RD48}:
\begin{eqnarray}
\label{eq:HamburgModel1}
N_{gl}(t)=&N_B \, (1-\eta(1-\exp(-c \cdot \Phi_{eq})))\,+\,N_{deep}(t) \\
\label{eq:HamburgModel2}
N_{deep}(t)=&N_a \,\exp(-t/\tau_a)+N_c +N_y\, (1-\exp(-t/\tau_Y)) \\ 
\label{eq:HamburgModel3}
&N_{a}=g_a \cdot \Phi_{eq} \quad N_c = g_c \cdot \Phi_{eq} \quad Ny=g_y \cdot \Phi_{eq} \quad ,
\end{eqnarray}
where $g_c$ is the introduction rate of defects constant in time ($N_c$), $g_a$ is the introduction
rate of defects that deactivate in time ($N_a$; short term annealing), $g_y$ the introduction
rate of defects that activate in time ($N_Y$; long term annealing), with time constants $\tau_a$ and $\tau_{Y}$.
The most critical parameter determining the properties of the gain layer is the acceptor removal constant $c(\Phi_{eq},t)$ which was found to be in the
range of $c \sim 5\cdot10^{-16}$ cm$^{2}$ for $N_B=10^{16}-10^{17}$ cm$^{-3}$, at which a complete removal  of acceptors can be assumed (fraction of initial acceptors removed $\eta=1$)
\cite{LGADRadHard}. 

The LGADs will be exposed up to equivalent fluences of $2.5\cdot 10^{15}$ cm$^{-2}$ at HL-LHC \cite{ATLAS,CMS}.
The upper limit to the contribution of deep acceptors ($(g_Y+g_C)\,\Phi_{eq}$) to the space charge in gain layer will
therefore be $\le 1.6 \cdot 10^{14}$ cm$^{-3}$, probably less for the negative deep acceptors can be neutralized by trapped holes
originating from the impact ionization. Therefore at any time the concentration of
initial acceptors will dominate the effective doping concentration in the gain layer.
The deep acceptors will however prevail in the bulk, for which the same equations \ref{eq:HamburgModel1}, \ref{eq:HamburgModel2}, \ref{eq:HamburgModel3} hold, but the
concentration of initial acceptors is several orders of magnitude smaller. The bulk doping determines the depletion
voltage of the device and the electric field in which carriers drift. The electric field is $E(x) \propto \int_{0}^x N_{eff}(x') dx'$,
therefore for a much thicker bulk than gain layer the electric field in the gain layer is affected by the bulk $N_{eff}$. This is particularly true at very large gains where a small change in electric field leads to substantial gain change.

The annealing will also affect the initial dopants ($N_B$) as the interstitial silicon atoms created by irradiation are mobile
and react with substitutional boron and deactivate it \cite{RW1995}. This process affects the removal
constant ($c$ in Eq. \ref{eq:HamburgModel1}) which becomes time dependent. A model for its observed fluence dependence
is proposed in Ref. \cite{CPhi}. 

The deep defects will on the other hand introduce generation current which will largely
originate in the bulk. The generation current decreases with annealing time and will lead to a reduction of leakage current at a given gain.

\section{Experimental setup and sample irradiations}

\begin{table}
\caption{
Devices used in the study. All the devices were single pads of $1.3\times1.3$ mm$^2$.
The low resistivity (0.01 $\Omega$cm) substrate was 200 $\mu$m thick and grown with
Czochralski technique. Back contacts were metallized.
The samples have opening in the metallization in the front contact allowing for light injection
and have two contacting regions for needle probing and bump bonding. The fluences in bold were used for timing measurements
only. The definition of the full depletion votlage $V_{fd}$ and the gain layer depletion votlage $V_{gl}$ are given later in the paper.
}
\label{ta:samples}
\begin{center}
\begin{tabular}{|c|c|c|c|c|c|}
\hline
Sample & Thickness & $V_{gl}$ & $V_{fd}$ & $\Phi_{eq} \quad [10^{14}]$ \\
\hline
HPK-1.1-35 & 35 $\mu$m  & 31 V & 195 V & 8, 15, 30 \\
HPK-1.2-35 & 35 $\mu$m  & 33 V & 36 V & 8, 15, 30  \\
HPK-3.1-50 & 50 $\mu$m  & 42 V & 49 V & 8, 15, 30  \\
HPK-3.2-50 & 50 $\mu$m  & 56 V & 64 V & {\bf 4}, 6, {\bf 8}, 15, {\bf 22.5}, 30  \\
\hline
\end{tabular}
\end{center}
\end{table}
The properties of the single pad LGADs from HPK \footnote{Hamamatsu Photonics, Japan} are listed in Table \ref{ta:samples}.
The samples were irradiated with reactor neutrons at TRIGA II research reactor of Jozef Stefan Institute \cite{Reaktor} to different
equivalent fluences. The annealing behavior of the samples was investigated using capacitance-voltage and current-voltage measurements (CV/IV) and Timing measurement with $^{90}$Sr electrons.

\begin{figure}[!hbt]
\begin{center}
\epsfig{file=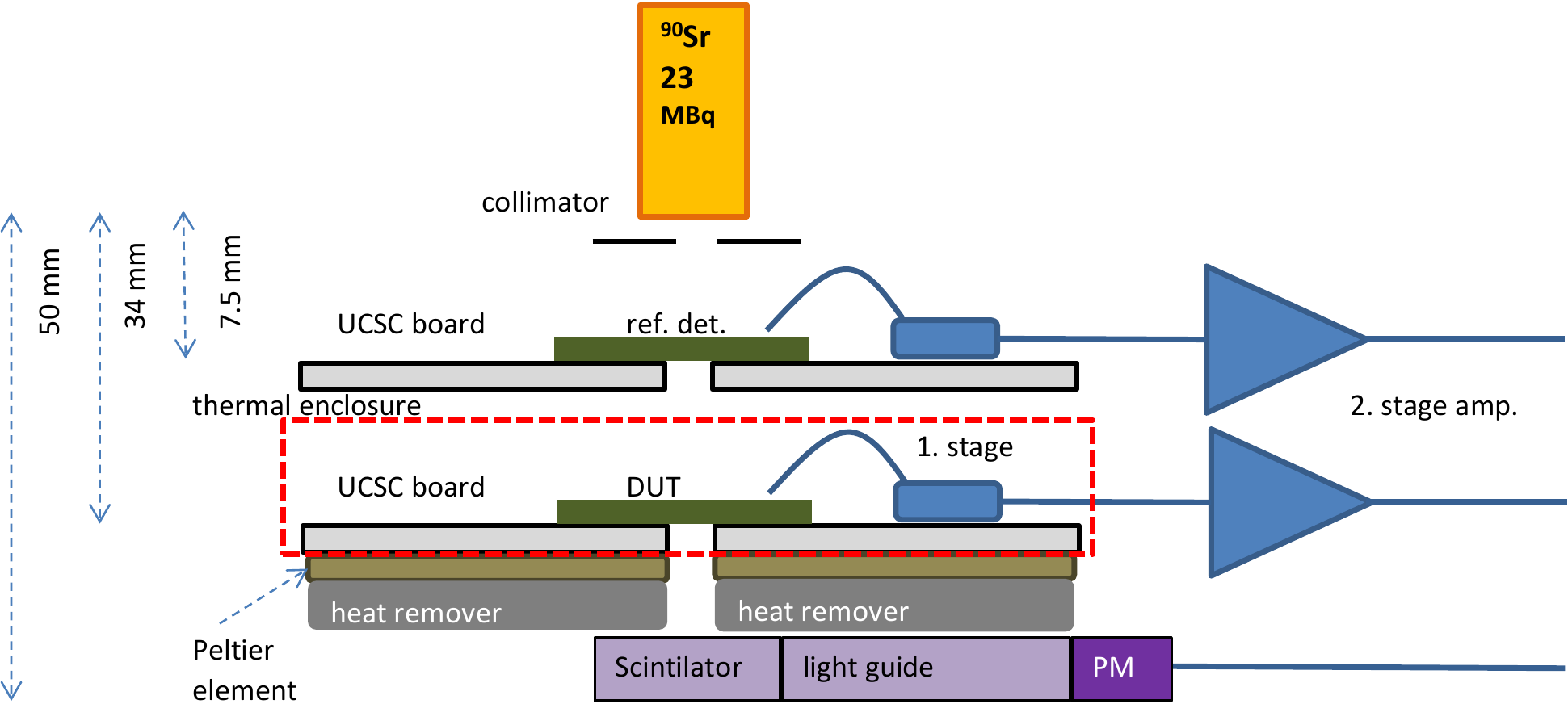,width=0.75\linewidth,clip=} \\
\end{center}
\caption{Experimental setup used for measurements of the time resolution.}
\label{fi:Setup}
\end{figure}

The timing measurements were conducted on 50 $\mu$m thick samples, HPK-3.1-50 and HPK-3.2-50, irradiated to several fluences.
The measurements were taken at $-30^\circ$C between the annealing steps at 60$^\circ$C.
The schematics of the measurement setup is shown in
Fig. \ref{fi:Setup}. The system consisted from a reference timing detector (50 $\mu$m thick LGAD from HPK, 0.8 mm in diameter
with gain of 60 \cite{GK2018,UCSC}) and small size scintillator
coupled with PM below it. The system was triggered by a coincidence of both detectors. The device under test (DUT) was
placed between these two detectors and carefully aligned to allow for a good fraction ($\sim 30-50\%$) of events
to hit all three sensors even for small pads of $\sim 1$ mm$^2$. The DUT was cooled with a Peltier and closed circle chiller.

The sensors were mounted
on electronics boards designed by UCSC \cite{UCSC}. The first stage of amplification uses a fast trans-impedance
amplifier (470 $\Omega$) followed by a second commercial amplifier (Particulars AM-02B,
35 dB, $>3$ GHz) which gives signals large enough to be recorded by a 40 GS/s digitizing
oscilloscope with 2.5 GHz bandwidth. 
\begin{figure}[!hbt]
\begin{tabular}{cc}
\epsfig{file=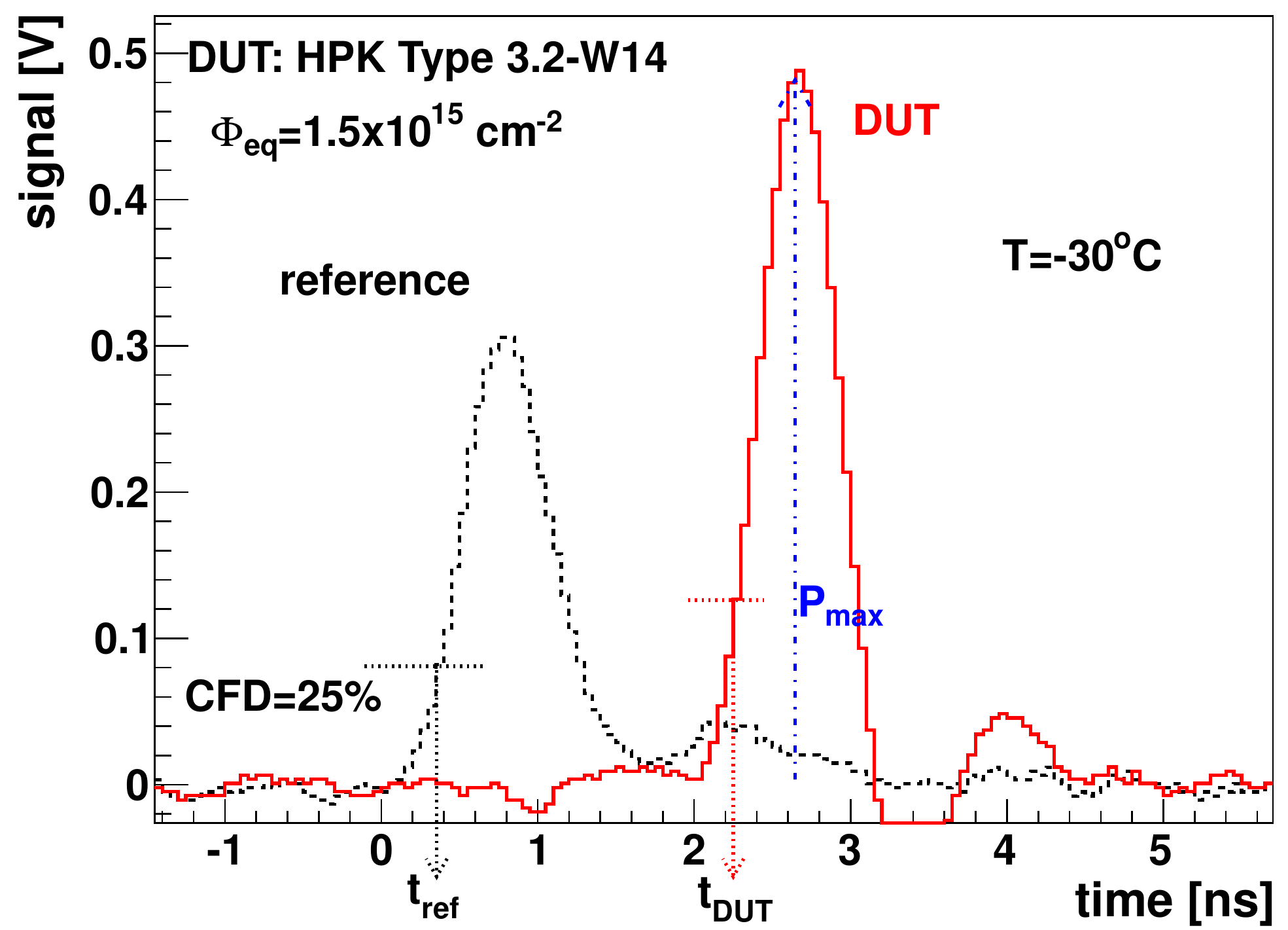,width=0.5\linewidth,clip=} & \epsfig{file=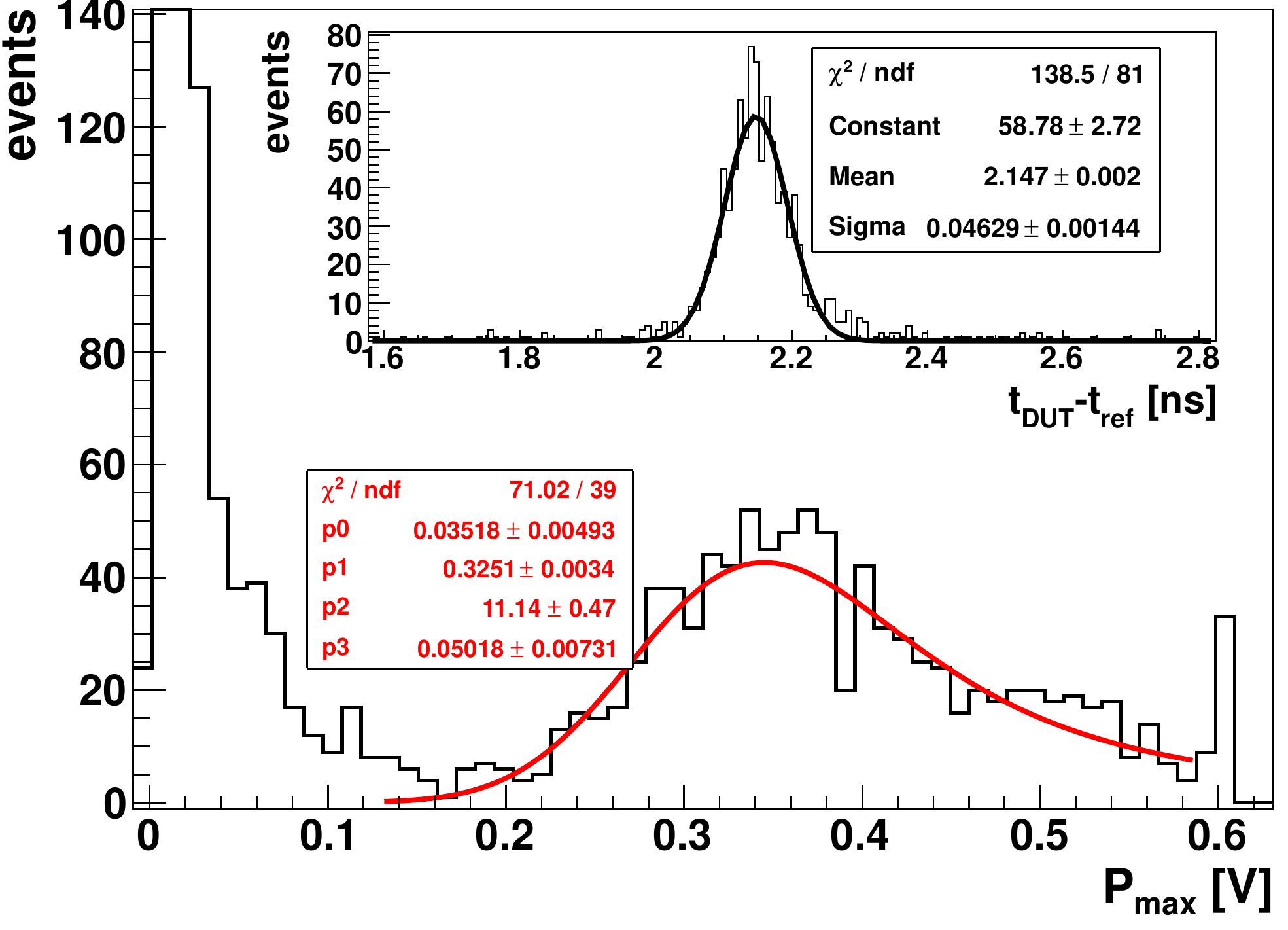,width=0.5\linewidth,clip=} \\
(a) & (b) 
\end{tabular}
\caption{(a) A typical signals recorded after electron passing reference and test detector. Constant fraction discrimination
at 25\% $P_{max}$ was used to determine hit times $t_{ref},t_{DUT}$. (b) Signal spectrum ($P_{max}$) with a fit of Landau-Gauss convolution. The
inset shows the spectrum of $t_{DUT}-t_{ref}$ and determination of time resolution from Gaussian fit to it. }
\label{fi:analysis}
\end{figure}

A typical event is shown in Fig. \ref{fi:analysis}a. The timing is determined from the time difference of both
signals crossing the threshold. A constant fraction discrimination was used, with the trigger occurring at the time
where the signal reached 25\% of its maximum. The maximum of the signal response was taken as the measure
of the charge. With fast amplifiers the integral of the current pulse is more commonly used, but for short pulses
in thin devices the difference is practically negligible. The spectrum of collected charge for a given device was
fitted with a convolution of Gaussian and Landau function and the most probable value was taken as collected charge
(see Fig. \ref{fi:analysis}b). The measured charge ($P_{max}$) was converted to absolute charge from unirradiated
sensor measurements in a calibrated system using slow electronics \cite{GK2018}.

The time resolution of the system was obtained as a standard deviation ($\sigma_{meas}$) of
the distribution $t_{DUT}-t_{ref}$. A typical distribution and a Gaussian fit to it is shown in
the inset of Fig. \ref{fi:analysis}b. The resolution of the reference detector was calibrated before and
was $\sigma_{ref}=30$ ps. The time resolution of the investigated detector is then obtained
as $\sigma_{DUT}^2=\sigma_{meas}^2-\sigma_{ref}^2$

\begin{figure}[!hbt]
\begin{tabular}{cc}
\epsfig{file=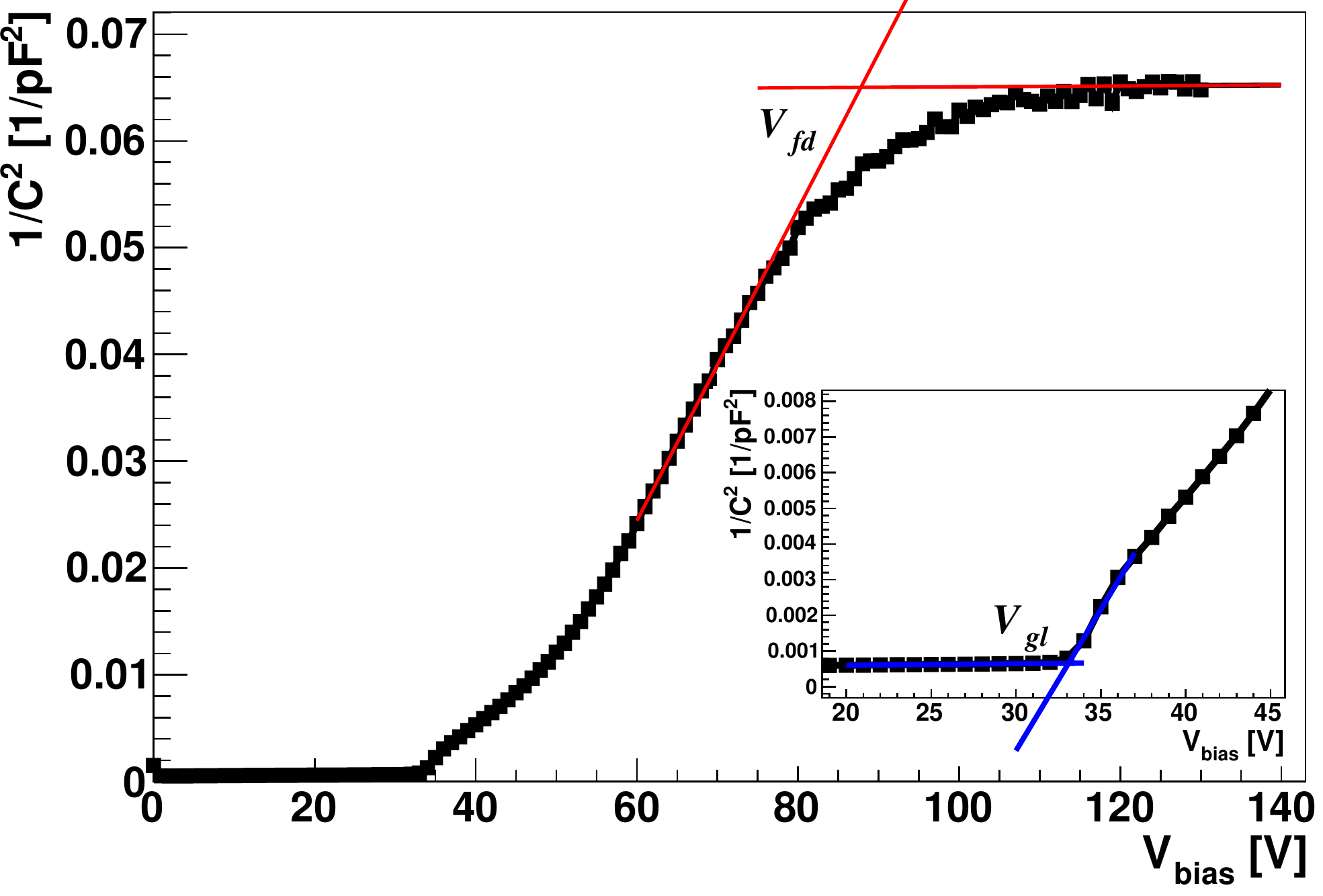,width=0.5\linewidth,clip=} & \epsfig{file=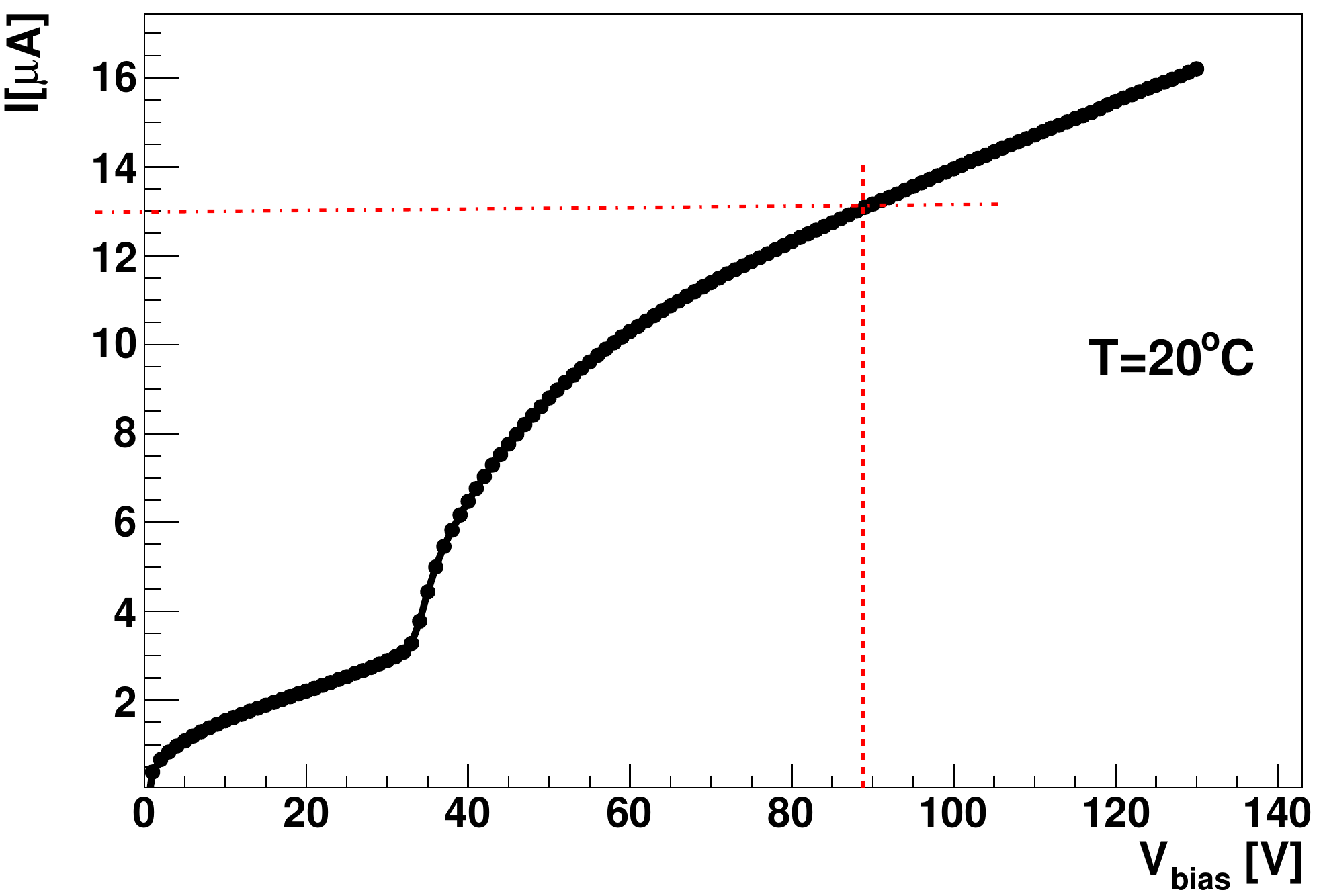,width=0.5\linewidth,clip=} \\
(a) & (b) 
\end{tabular}
\caption{ (b) An example of Capacitance-Voltage measurements for HPK-3.2-50 device irradiated to $\Phi_{eq}=1.5\cdot 10^{15}$ cm$^{-2}$ with determination of 
$V_{fd}$ and $V_{gl}$ (in inset). (b) Determination of the generation current as current measured at $V_{fd}$.}
\label{fi:CVIV-Measurements}
\end{figure}

CV/IV measurements were taken in the probe station at $T=20^\circ$C and $\nu=10$ kHz using
Keithley 6517 for voltage source and precise current meter and LCR meter HP4263B. The guard ring was
grounded during the measurements. An example of $V_{fd}$ and $V_{gl}$ extraction is shown
in Fig. \ref{fi:CVIV-Measurements}a and a corresponding leakage current
at $V_{fD}$ in Fig. \ref{fi:CVIV-Measurements}b.

\section{Annealing of the gain layer doping concentration}
\begin{figure}[!hbt]
\begin{tabular}{cc}
\epsfig{file=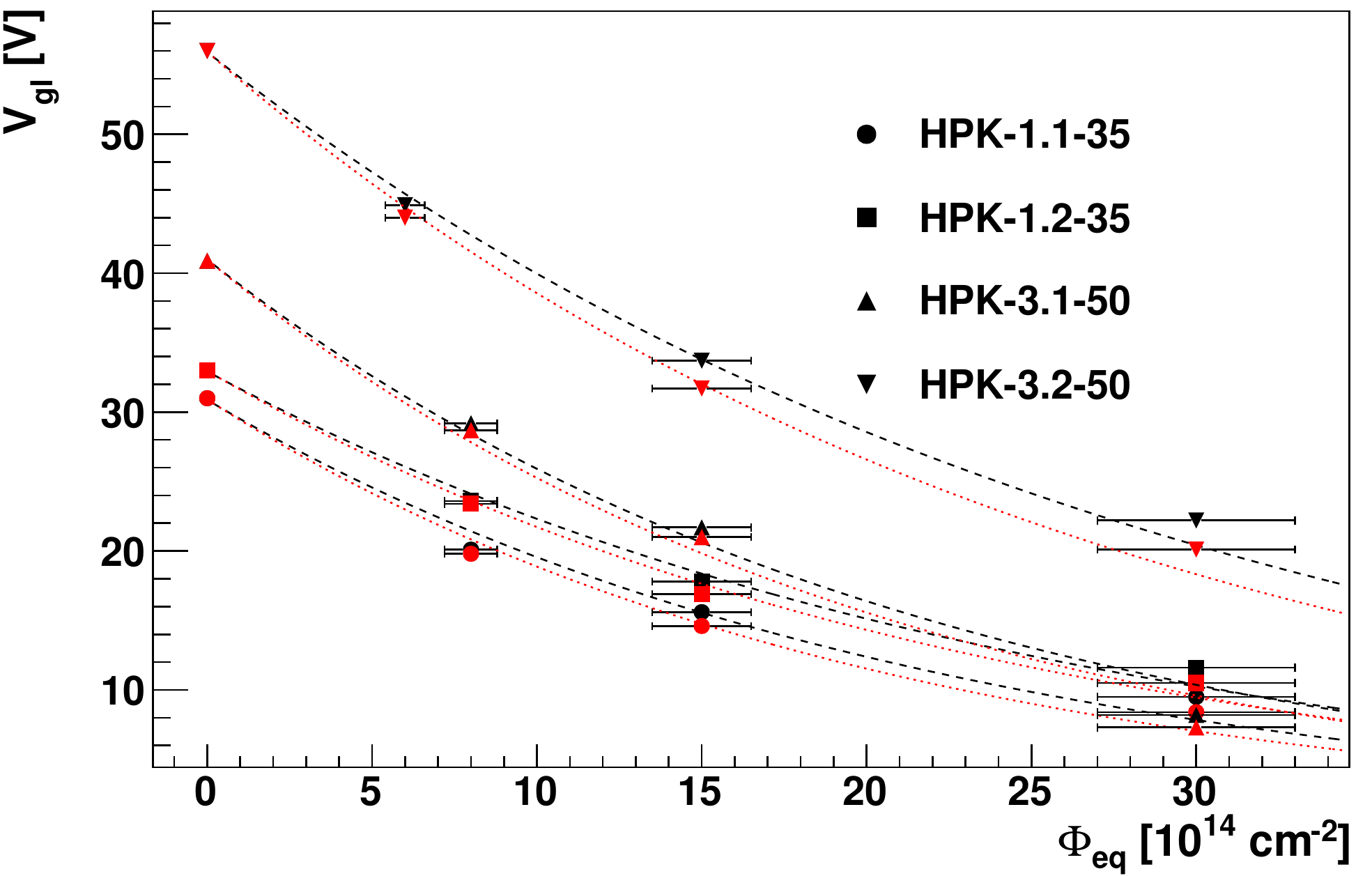,width=0.5\linewidth,clip=} & \epsfig{file=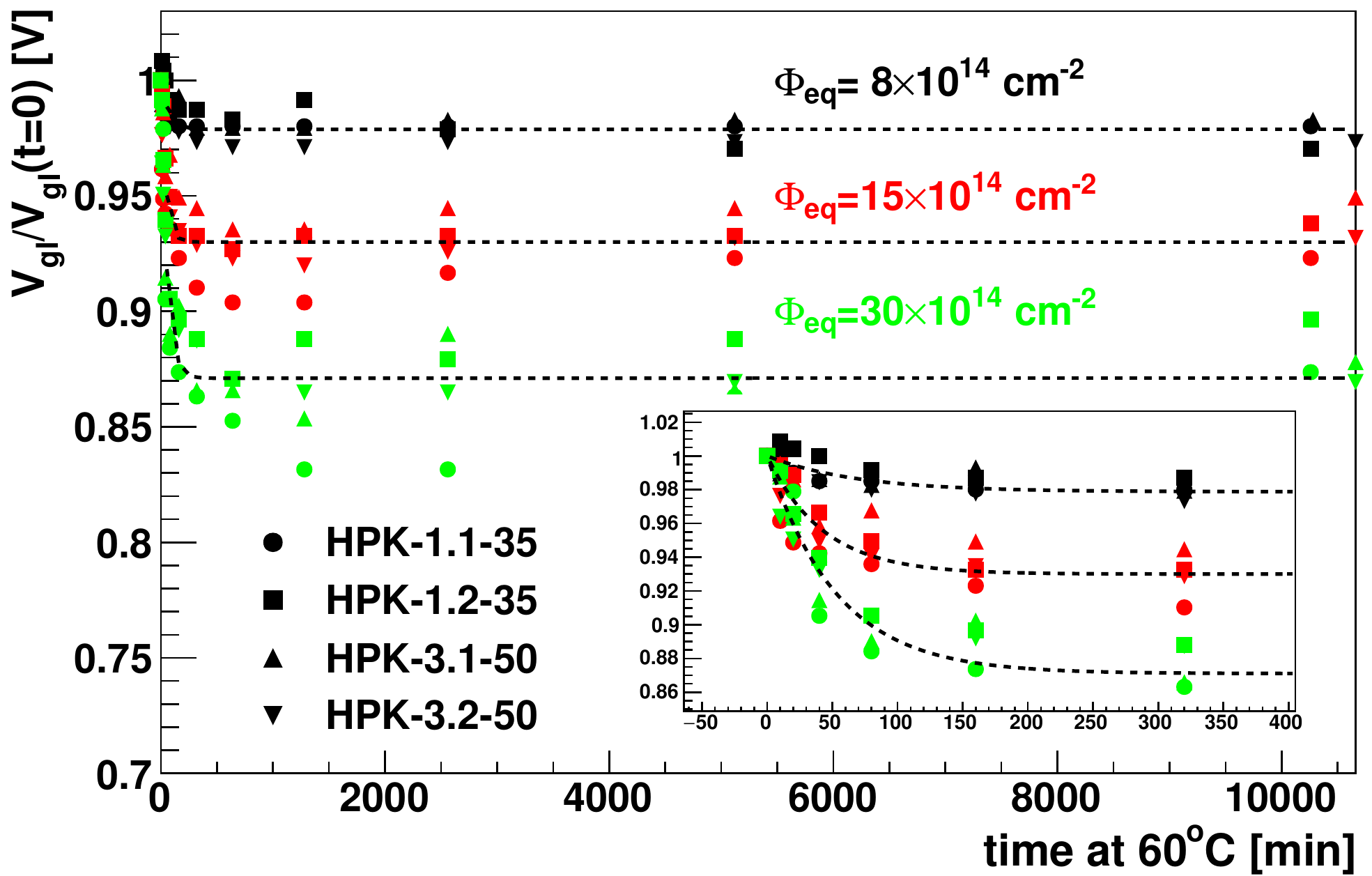,width=0.5\linewidth,clip=} \\
(a) & (b) 
\end{tabular}
\caption{(a) Dependence of $V_{gl}$ on fluence for investigated sensors
before any intentional annealing (black) and after 80 min annealing (red). (b) Relative change of $V_{gl}$ with annealing for different samples and fluences. The marker shape denotes the sensor type and the color the fluence. The inset shows shorter time scale. The global
 fit of Eq. \ref{eq:VglTime} is shown by the dashed line.}
\label{fi:VglFluTime}
\end{figure}

The dependence of $V_{gl}$ on fluence immediately after irradiation (0 min) and after 80 min of annealing at 60$^\circ$C is
shown in Fig. \ref{fi:VglFluTime}a. Assuming that $V_{gl}$ is proportional to the $N_{gl}$, its dependence on fluence can be parameterized as
\begin{equation}
V_{gl}(\Phi_{eq},t)\,=\,V_{gl}(0)\cdot \exp(-c(t)\cdot\Phi_{eq}) \quad .
\label{eq:VglFlu}
\end{equation}
The fits of Eq. \ref{eq:VglFlu} to the measured data are also shown and the obtained removal constants are gathered in Table \ref{ta:VGL-Fluence}. Annealing impacts the gain layer in a moderate way, but even small changes can lead to a substantial impact in gain operation mode as will be discussed later.

The difference in removal constants for different samples can be attributed to the different doping
levels of the various samples and to the shapes of the doping profiles. This information is not
revealed by the producer, but the effective doping profiles extracted from C-V reveal a deeper p$^+$ layer
for HPK-3.2-50 than for HPK-3.1-50.
\begin{table}
\caption{Parameters obtained from the fit of Eq. \ref{eq:VglFlu} to the data shown in Fig. \ref{fi:VglFluTime}.}
\label{ta:VGL-Fluence}
\begin{center}
\begin{tabular}{|c|c|c|c|c|c|}
\hline
  & HPK-1.1-35 & HPK-1.2-35 & HPK-3.1-35 & HPK-3.2-35  \\
\hline
$c(t=0) [10^{-16}$ cm$^{-2}]$                     & 4.5$\pm$0.3 & 3.9$\pm$0.3 & 4.5$\pm$0.3 & 3.3$\pm$0.2  \\
$c(t=80\,\,\mathrm{min}) [10^{-16}$ cm$^{-2}]$   & 4.9$\pm$0.3 & 4.2$\pm$0.3 & 4.8$\pm$0.3 & 3.7$\pm$0.3  \\
\hline
\end{tabular}
\end{center}
\end{table}

The $V_{gl}$ depends on time after irradiation. A change relative to
$V_{gl}(t=0)$ is shown in  Fig. \ref{fi:VglFluTime}b. It seems that the relative change becomes larger with
fluence, but doesn't depend on the sample type. The data for all samples at a given fluence 
were therefore fitted with
\begin{equation}
\frac{V_{gl}}{V_{gl}(t=0)}=F \cdot \exp(-t/\tau_{gl}) + (1-F) \quad ,
\label{eq:VglTime}
\end{equation}
where $F$ represents the fraction of effective dopants that change in time and $\tau_{gl}$ the time
constant for the process. The results of the global fit with $F$ and $\tau_{gl}$ to the all the data
at a given fluence are shown in Table \ref{ta:VGL-time}. The increase of the $F$ with fluence is evident
and amounts to a change in $V_{gl}$ of more than 2 V at $\Phi_{eq}=1.5\cdot 10^{15}$ cm$^{-2}$ for HPK-3.2-50. This has an important
impact on device performance as will be shown later.  
\begin{table}
\caption{Parameters obtained from the fit of Eq. \ref{eq:VglTime} to the data shown in Fig. \ref{fi:VglFluTime}b. The
fit uncertainties are obtained assuming 0.4 V uncertainty in $V_{gl}$ determination from C-V.}
\label{ta:VGL-time}
\begin{center}
\begin{tabular}{|c|c|c|c|}
\hline
$\Phi_{eq} [10^{14}$ cm$^{-2}]$  & 8 & 15 & 30 \\
\hline
$F$               & $0.021\pm 002$ & $0.07\pm 0.003$ & $0.13 \pm 0.003$ \\
$\tau_{gl}$ [min]  & $68 \pm 19$ & $43 \pm 8$ & $53 \pm 6$  \\
\hline
\end{tabular}
\end{center}
\end{table}
The time constants for different fluences $\tau_{gl}$ were found to be compatible
with an average of $\tau_{gl}=50 \pm 5$ min.

\begin{figure}[!hbt]
\begin{center}
\epsfig{file=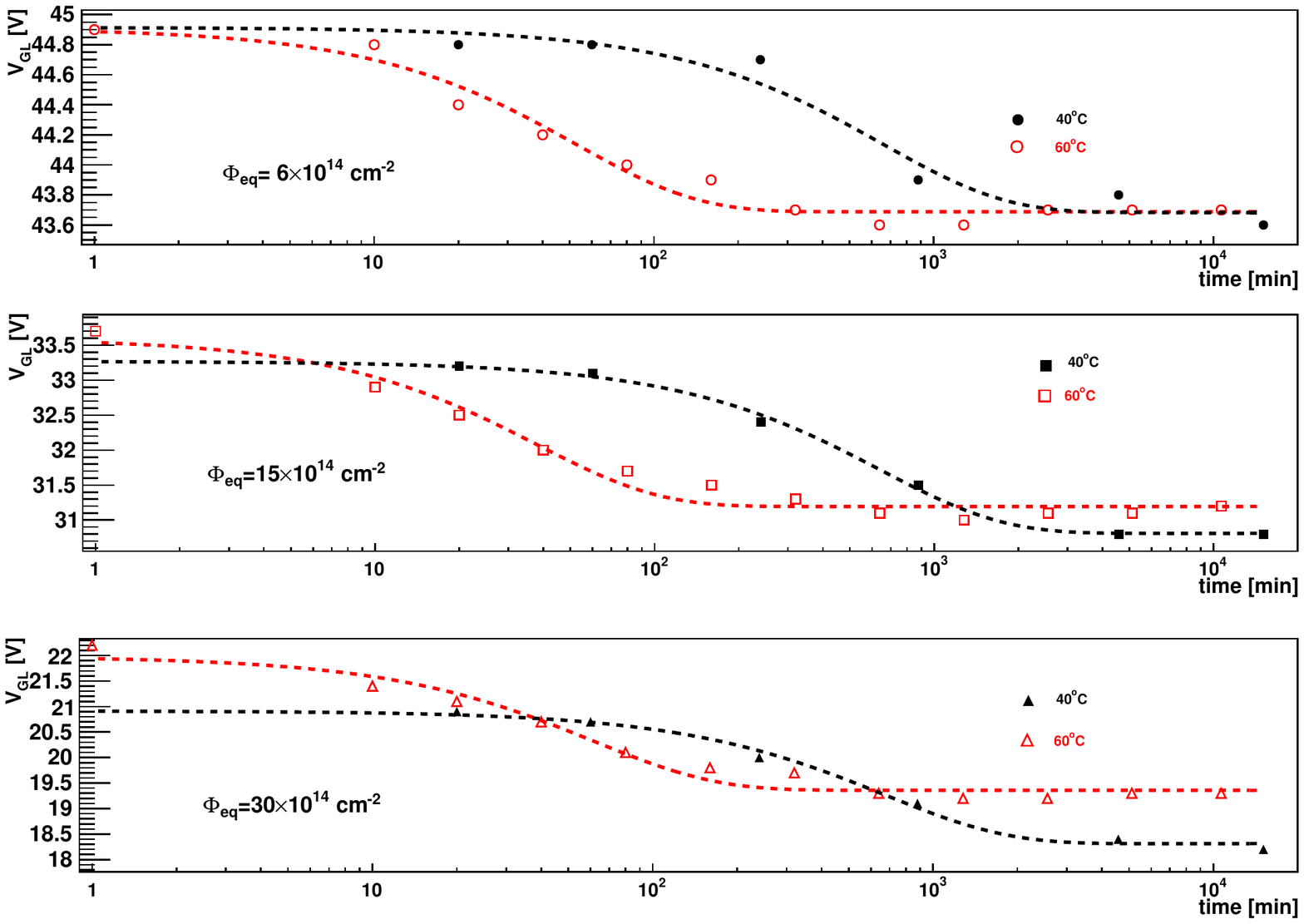,width=0.7\linewidth,clip=} \\
\end{center}
\caption{Evolution of $V_{gl}$ with annealing time at 40$^\circ$C (solid markers) and 60$^\circ$C (open markers)
for different fluences.  }
\label{fi:VGL-diffT}
\end{figure}

In order to determine the dynamics of these changes at lower temperatures a set of sensors of type HPK-3.2-50 were
also annealed at $T=40^\circ$C, which enabled the extraction of the activation energy for the process and thus allow scaling to
temperatures close to operation temperatures. The comparison of $V_{gl}$ annealing for $40^\circ$C and $60^\circ$C is shown
in Fig. \ref{fi:VGL-diffT}. The fit of Eq. \ref{eq:VglTime} for $V_{gl}(t)$ to the measured data is also shown.
The time constants for all three fluences were found comparable for each temperature, yielding an
average over the fluences of $\tau_{gl}(40^\circ\mathrm{C})=652 \pm 50$ min and $\tau_{gl}(60^\circ\mathrm{C})=47 \pm 6$.
The fraction of $V_{gl}$ annealed seems to be the same for both temperatures and is in accordance with the $F$ given in Table \ref{ta:VGL-time}. It is therefore reasonable to assume that $F$ would be the same also at other annealing temperatures.
As a process governed by thermal energy the Arrhenius relation can be used to determine the
activation energy ($E_a$)
\begin{equation}
\frac{\tau_{gl}(T_1)}{\tau_{gl}(T_2)}=\frac{\exp(\frac{E_a}{k_BT_1})}{\exp(\frac{E_a} {k_BT_2} )} \rightarrow E_a=\frac{k_B T_1 T_2}{T_2-T_1}\ln(\frac{\tau_{gl}(T_1)}{\tau_{gl}(T_2)})\quad ,
\label{eq:ArrTau}
\end{equation}
where $k_B$ is the Boltzmann constant. The activation energy $E_a=1.15 \pm0.07$ eV was obtained. Although uncertainties
are large, it allows the scaling to operational temperature of LGADs at HL-LHC $\tau_{gl}(\mathrm{-30^\circ C})=5.3$ years,
so annealing is effectively frozen during the yearly operation period.

\section{Annealing of the bulk doping concentration and generation current}
\subsection{Bulk doping concentration}
The evolution of the effective doping concentration in the bulk is usually dominated by the deep defects, as the high resistivity
silicon is normally used. Apart from the influence of the bulk on the electric field, the bulk generation
current is the major source of the leakage current. The effective doping concentration in the bulk was measured as
\begin{equation}
N_{eff}=\frac{ 2 \varepsilon \varepsilon_0 (V_{fd}-V_{gl})}{\mathrm{e_0}\,\,(D-x_{gl})^2} \quad,
\label{eq:bulkNeff}
\end{equation}
where $\varepsilon$ is permitivity of Si, $\varepsilon_0$ permittivity of vacuum, $\mathrm{e_0}$ elementary charge, $D$ active
thickness of the device and $x_{gl}$ the width of the gain layer. The latter is not precisely known, but is
around 2 $\mu$m. The choice of $x_{gl}$ has a marginal impact on $N_{eff}$. Eq. \ref{eq:bulkNeff} implies a
constant space charge, an assumption that proved fairly accurate for neutron irradiated sensors over past decades. The small
thickness of the devices allowed for accurate determination of $V_{fd}$ at relatively large fluences, which is not possible for
thicker sensors. On the other hand, mobile impurities, mostly oxygen, can migrate from the substrate and influence the
properties of the bulk. For the investigated fluences the decrease in $V_{gl}$ was such that, except for HPK-3.2-50,
there was no gain at full depletion of the detector, which only appeared at $V_{bias} \gg V_{fd}$.
Therefore, at voltages close to $V_{fd}$ there was no enhanced hole concentration in the bulk originating from the gain layer.
\begin{figure}[!hbt]
\begin{tabular}{cc}
\epsfig{file=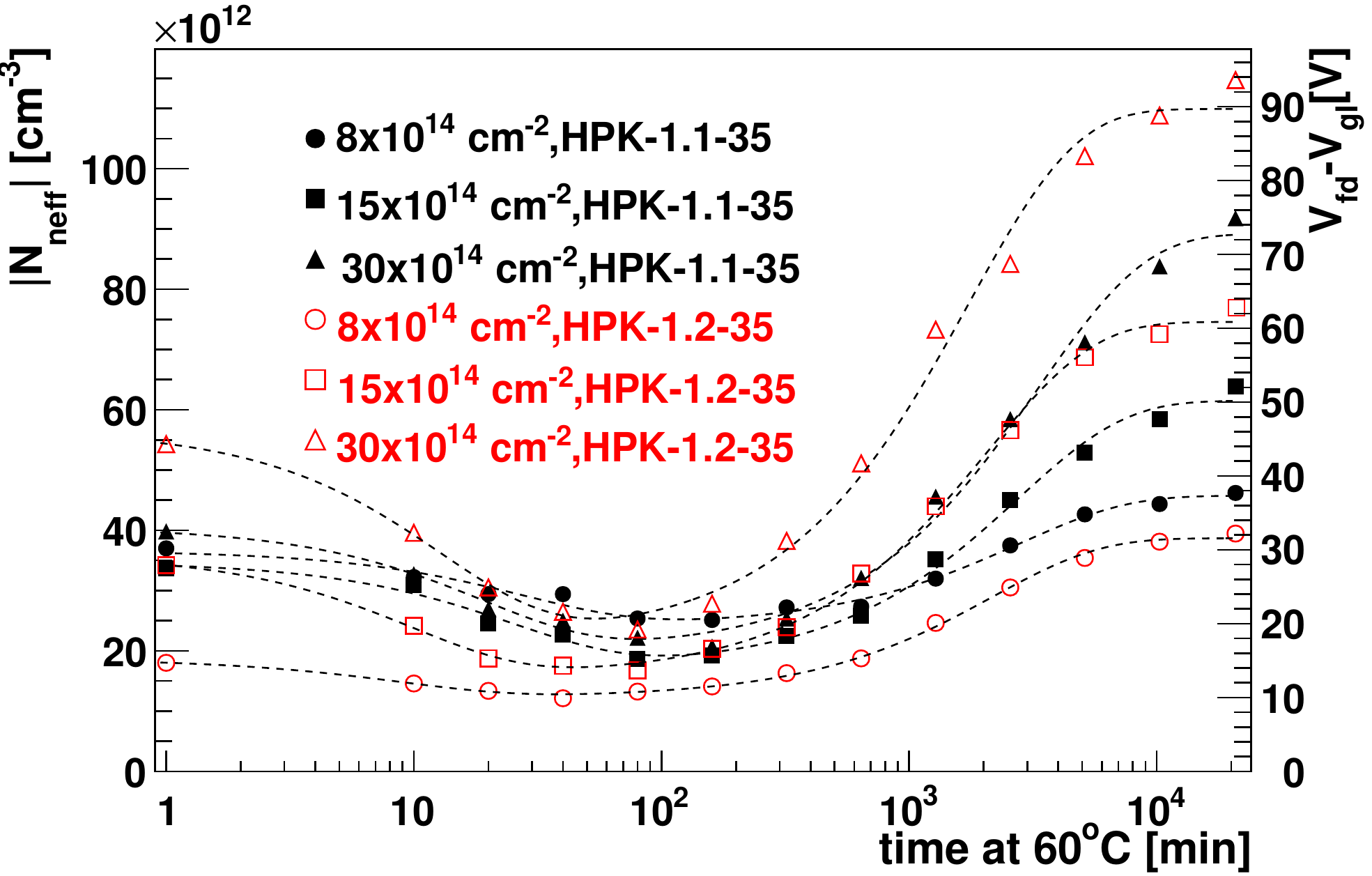,width=0.5\linewidth,clip=} & \epsfig{file=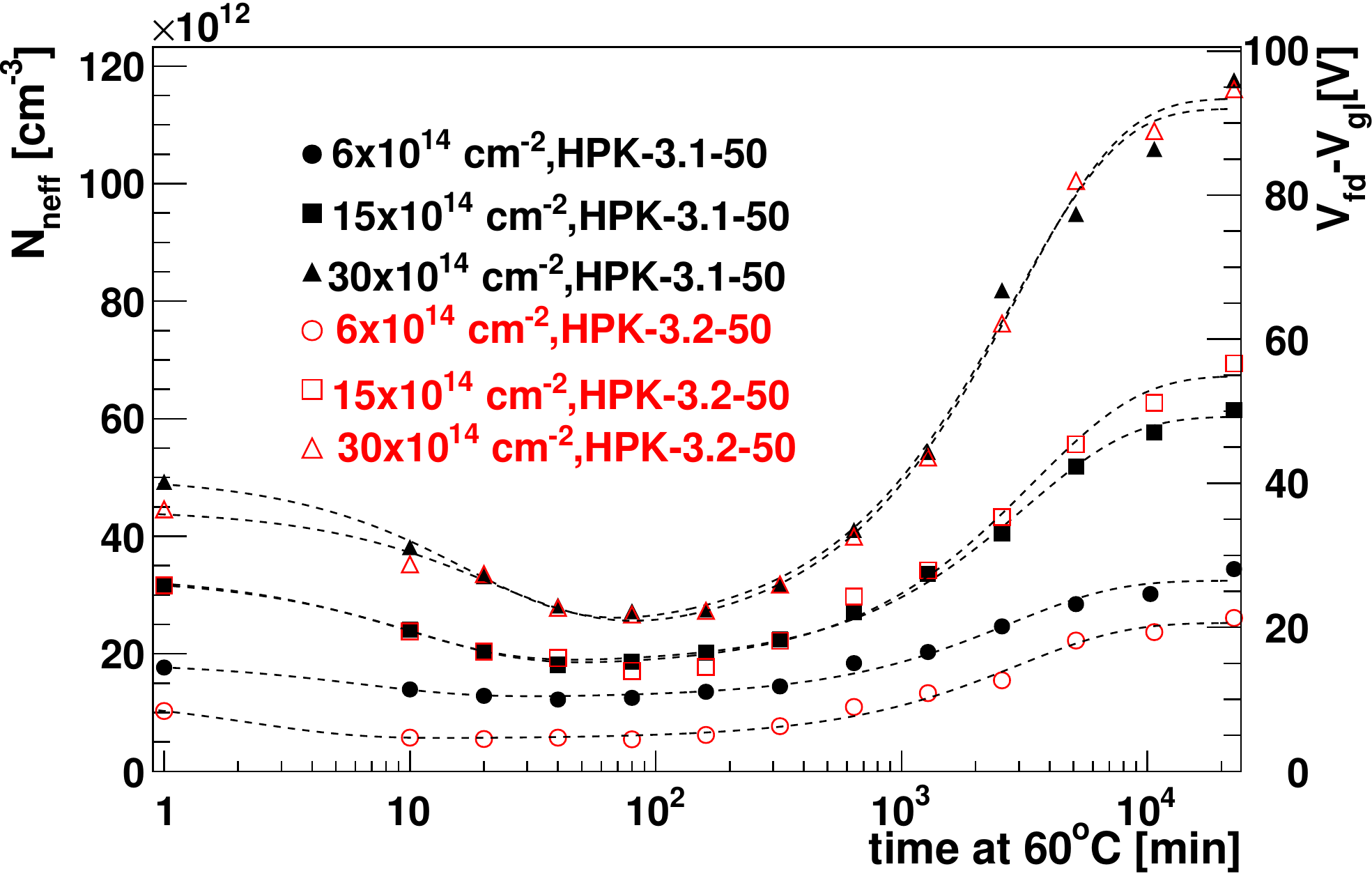,width=0.5\linewidth,clip=} \\
(a) & (b) 
\end{tabular}
\caption{Evolution of $N_{eff}$ with time after irradiation for (a) 35 $\mu$m and (b) 50 $\mu$m thick sensors. The fits of
Eqs. \ref{eq:HamburgModel2},\ref{eq:HamburgModel3} are also shown. Note that the point at $t=0$ min is shown at 1 min due to logarithmic scale. }
\label{fi:VfdTime}
\end{figure}

The evolution of the $N_{eff}$ for all the investigated sensors is shown in Figs. \ref{fi:VfdTime}. The sensors differ in initial bulk
resistivity which can be clearly observed in the $N_{eff}$ evolution. The initial boron removal is not yet completed for
HPK-1.1-35 at $8\cdot 10^{14}$ cm$^{-2}$ (Fig. \ref{fi:VfdTime}a), hence its minimum $N_{eff}$ in time is larger than that of the
other two fluences. The fits of the Hamburg model (Eqs. \ref{eq:HamburgModel2},\ref{eq:HamburgModel3}) to the all the measured
results are also shown and a good description can be observed. The model parameters for the samples in Table \ref{ta:samples} are listed in Table \ref{ta:VFD-time}.
\begin{table}
\caption{Parameters obtained from the fits to the data shown in Fig. \ref{fi:VfdTime}. The
fit uncertainties are obtained assuming 0.4 V uncertainty in $V_{gl}$ and 3 V in $V_{fd}$ determination from C-V.}
\label{ta:VFD-time}
\begin{center}
\begin{tabular}{|c|c|c|c|c|c|c|}
\hline
& $\Phi_{eq} [$ cm$^{-2}]$  & $N_C$ [$10^{13}$ cm$^{-3}$] & $g_a [10^{-2}$ cm$^{-1}]$ & $\tau_a$ [min] & $g_Y [10^{-2}$ cm$^{-1}$] & $\tau_{ra}$ [min]  \\
\hline
HPK-1.1-35 & 8$\cdot10^{14}$  & $2.41 \pm 0.1$ &  $1.57\pm0.2$ & $29 \pm 8$  & $2.4\pm 0.2$ & $2780\pm 400$ \\
HPK-1.1-35 & 15$\cdot10^{14}$ & $1.72 \pm 0.2$ &  $1.17\pm0.2$ & $27 \pm 7$  & $2.9\pm 0.2$ & $2740\pm 310$ \\
HPK-1.1-35 & 30$\cdot10^{14}$ & $2.00 \pm 0.2$ &  $0.7 \pm0.2$ & $21 \pm 7$  & $2.3\pm 0.2$ & $3330\pm 370$ \\
\hline
HPK-1.2-35 & 8$\cdot10^{14}$   & $1.22 \pm 0.1$ &  $0.81 \pm0.2$ & $9.4 \pm 3$  & $3.3\pm 0.1$ & $2170\pm 120$ \\
HPK-1.2-35 & 15$\cdot10^{14}$  & $1.57 \pm 0.1$ &  $1.36 \pm0.2$ & $10.4 \pm 2$  & $3.9\pm 0.2$ & $2060\pm 130$ \\
HPK-1.2-35 & 30$\cdot10^{14}$  & $2.2 \pm 0.2$ &  $1.16 \pm0.2$ & $13.7 \pm 5$  & $2.9\pm 0.2$ & $1740\pm 220$ \\
\hline
HPK-3.1-50 & 8$\cdot10^{14}$  & $1.25 \pm 0.2$ &  $0.75 \pm0.1$ & $6.5 \pm 3$  & $2.5\pm 0.1$ & $2750 \pm 520$ \\
HPK-3.1-50 & 15$\cdot10^{14}$  & $1.82 \pm 0.1$ &  $0.93 \pm0.1$ & $10 \pm 2$  & $2.8\pm 0.2$ & $3180 \pm 220$ \\
HPK-3.1-50 & 30$\cdot10^{14}$  & $2.35 \pm 0.1$ &  $0.89 \pm0.1$ & $18 \pm 6$  & $3.0\pm 0.2$ & $2840 \pm 320$ \\
\hline
HPK-3.2-50 & 6$\cdot10^{14}$   & $0.55 \pm 0.05$ &  $1.1 \pm0.2$ & $3 \pm 5$  & $3.1\pm 0.33$ & $2967 \pm 400$ \\
HPK-3.2-50 & 15$\cdot10^{14}$  & $1.77 \pm 0.01$ &  $1.05 \pm0.1$ & $12 \pm 4$  & $3.1\pm 0.33$ & $3415 \pm 420$ \\
HPK-3.2-50 & 30$\cdot10^{14}$  & $2.25 \pm 0.01$ &  $0.8  \pm0.1$ & $24 \pm 6$  & $3.0\pm 0.31$ & $2940 \pm 180$ \\
\hline
\end{tabular}
\end{center}
\end{table}

\begin{figure}[!hbt]
\begin{center}
\epsfig{file=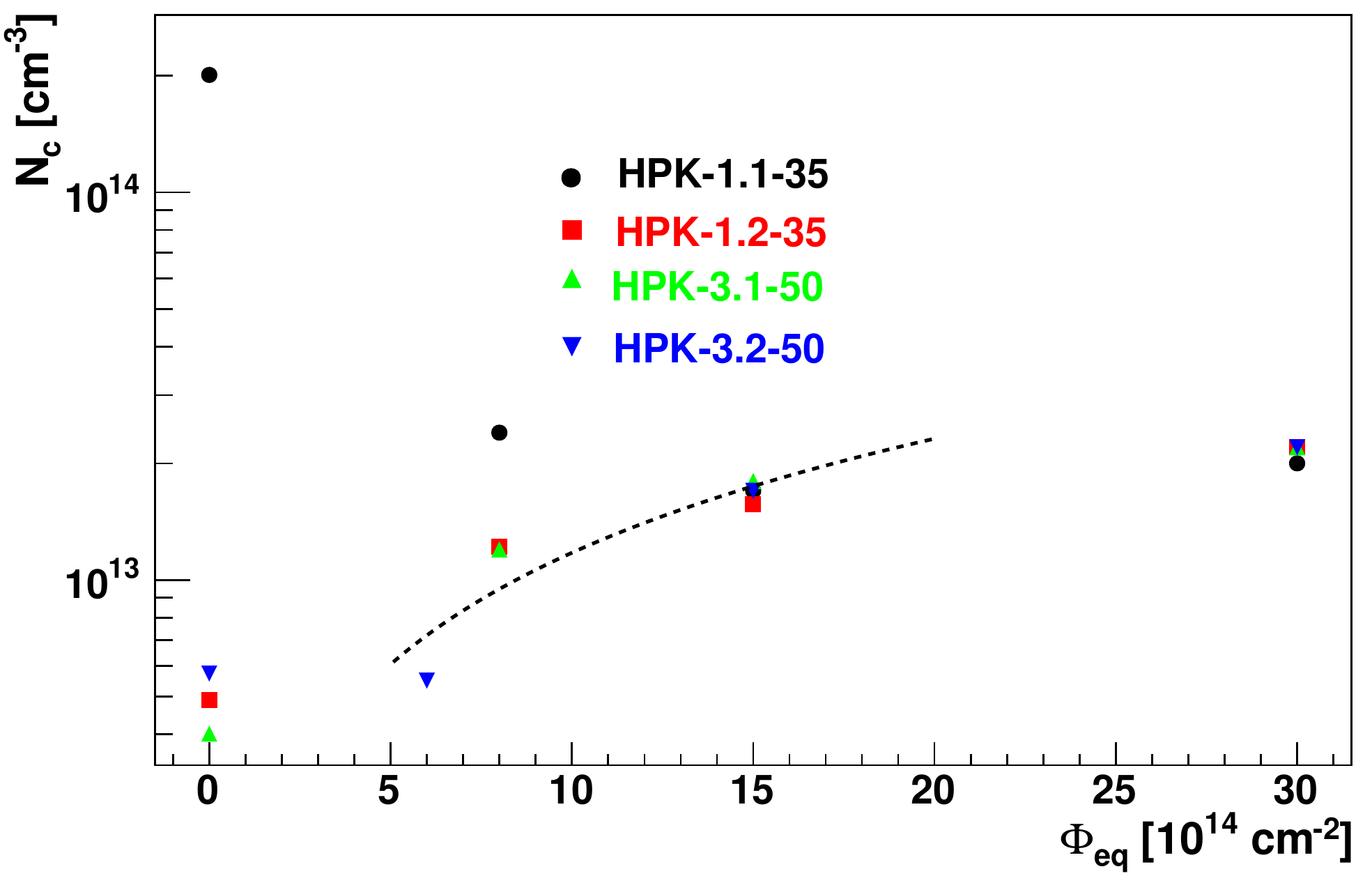,width=0.5\linewidth,clip=}  
\end{center}
\caption{Dependence of stable acceptors concentration on fluence. The fit to the data excluding HPK-1.1-35 and assuming
full acceptor removal in the bulk is also shown. The highest fluence point was excluded as explained in the text.}
\label{fi:NcVsFlu}
\end{figure}

The dependence of stable damage on equivalent fluence is shown in Fig. \ref{fi:NcVsFlu}. Apart
from HPK-1.1-35, the resistivity of samples was high enough for the initial acceptor concentration
to be less important. Assuming that the acceptors were completely removed for those samples by irradiation down to the lowest
fluence, and excluding the highest fluence point where the saturation of stable damage was earlier observed \cite{GK2014},
the slope $dN_C/d\Phi_{eq}$ showed the result to be $g_c=0.012 \pm 0.002$ cm$^{-1}$, which is somewhat lower
than $g_c=0.017 \pm 0.2$ cm$^{-1}$ measured for detectors at standard thickness irradiated to fluences $<10^{15}$ cm$^{-2}$ \cite{RD48}. 

It has been observed before that thin sensors can have different damage parameters \cite{GL2005} also depending on
properties of the support wafer. Introduction of defects that reverse anneal ($g_y$) were found
to be smaller, while the time constants ($\tau_{ra}\sim 2500-3500$ min) were found to be longer than in standard FZ silicon, but both roughly compatible with those in oxygenated silicon detectors. 
The short term annealing results are compatible with previous measurements in standard silicon detectors
with an introduction rate of $g_a \sim 0.1$ cm$^{-1}$ and annealing times of $\tau_a \approx 20$ min, but the uncertainty is
rather large. It is worth mentioning that at low $V_{fd}$ the relative uncertainty in $V_{fd}$ determination becomes larger. 

At bias voltages high enough for charge multiplication 
the free hole concentration in the bulk increases which decreases the negative space charge in irradiated detectors. The
dependence of traps' occupancy on free hole concentration was
studied by free hole injection through continuous illumination of the surface with red light \cite{GK2003}. 
The concentration
of free holes in the LGAD bulk can be estimated from the leakage current ($I_{leak}$). As will be shown
in the next section currents of few $\mu$A are measured at high bias voltages and -30$^\circ$C. Assuming
that holes contribute dominantly to the  leakage current for high gain $G$ (hole contribution $\sim G/(G+1)$) the free hole
concentration ($p$) is given by
\begin{equation}
p \approx \frac {I_{leak}}{\mathrm{e_0}\, v_{sat,p}\,S} \quad,
\end{equation}
where $v_{sat,p}=6\cdot10^6$ cm/s is saturation velocity of the holes and $S=0.017$ cm$^2$ is the surface of the detector. The concentration of free holes is therefore in the order of $p \approx 2\cdot 10^8$ cm$^{-3}$ for $I_{leak}=3$ $\mu$A. Such a concentration influences the space charge, but it still remains negative in the bulk \cite{GK2003}.
The effective doping concentration calculated with Eqs. \ref{eq:HamburgModel2}, \ref{eq:HamburgModel3}
and parameters from Table \ref{ta:VFD-time} set the upper limit for the negative space charge inside the detector bulk.
\begin{figure}[!hbt]
\begin{tabular}{cc}
\epsfig{file=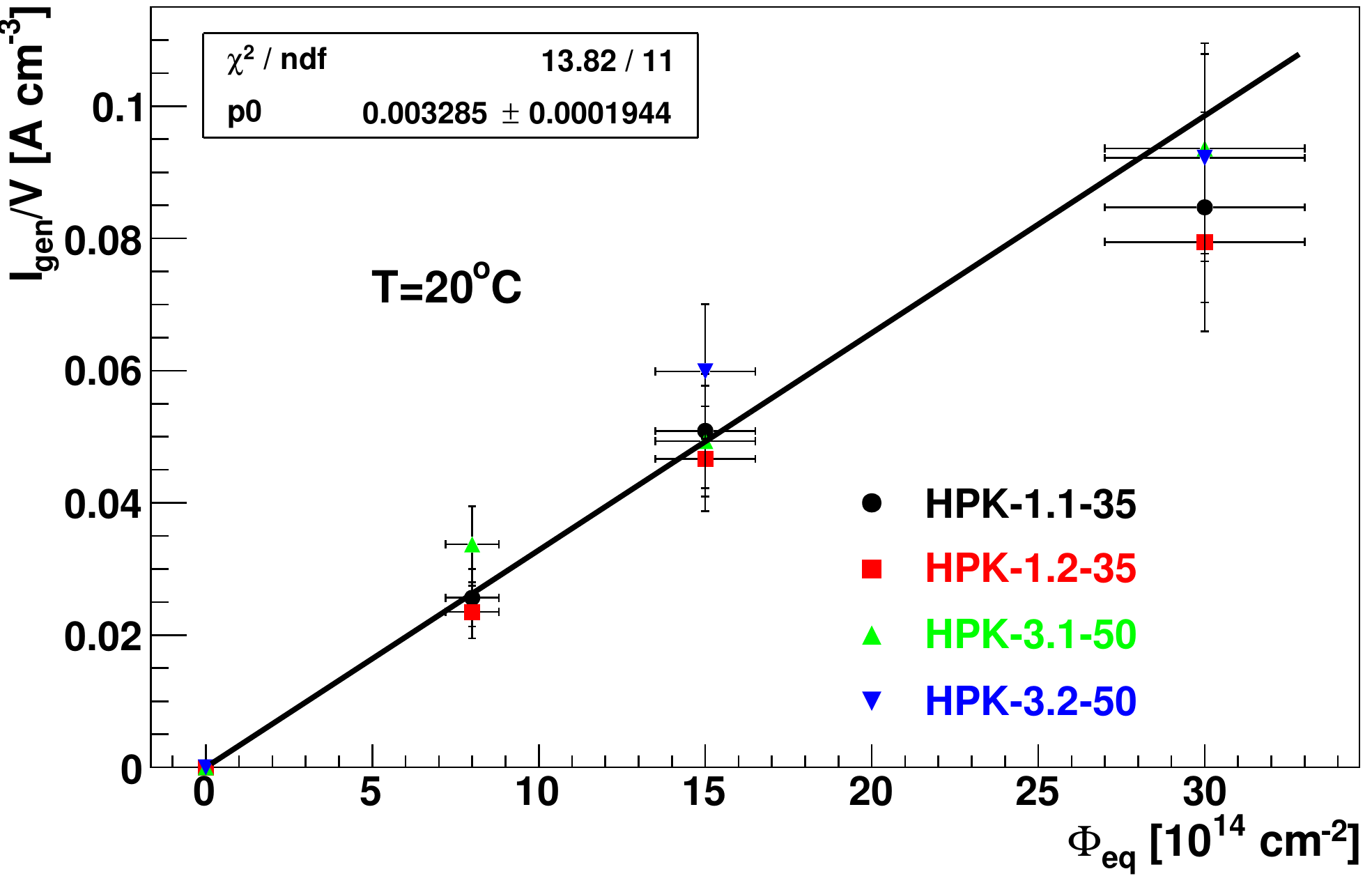,width=0.5\linewidth,clip=} & \epsfig{file=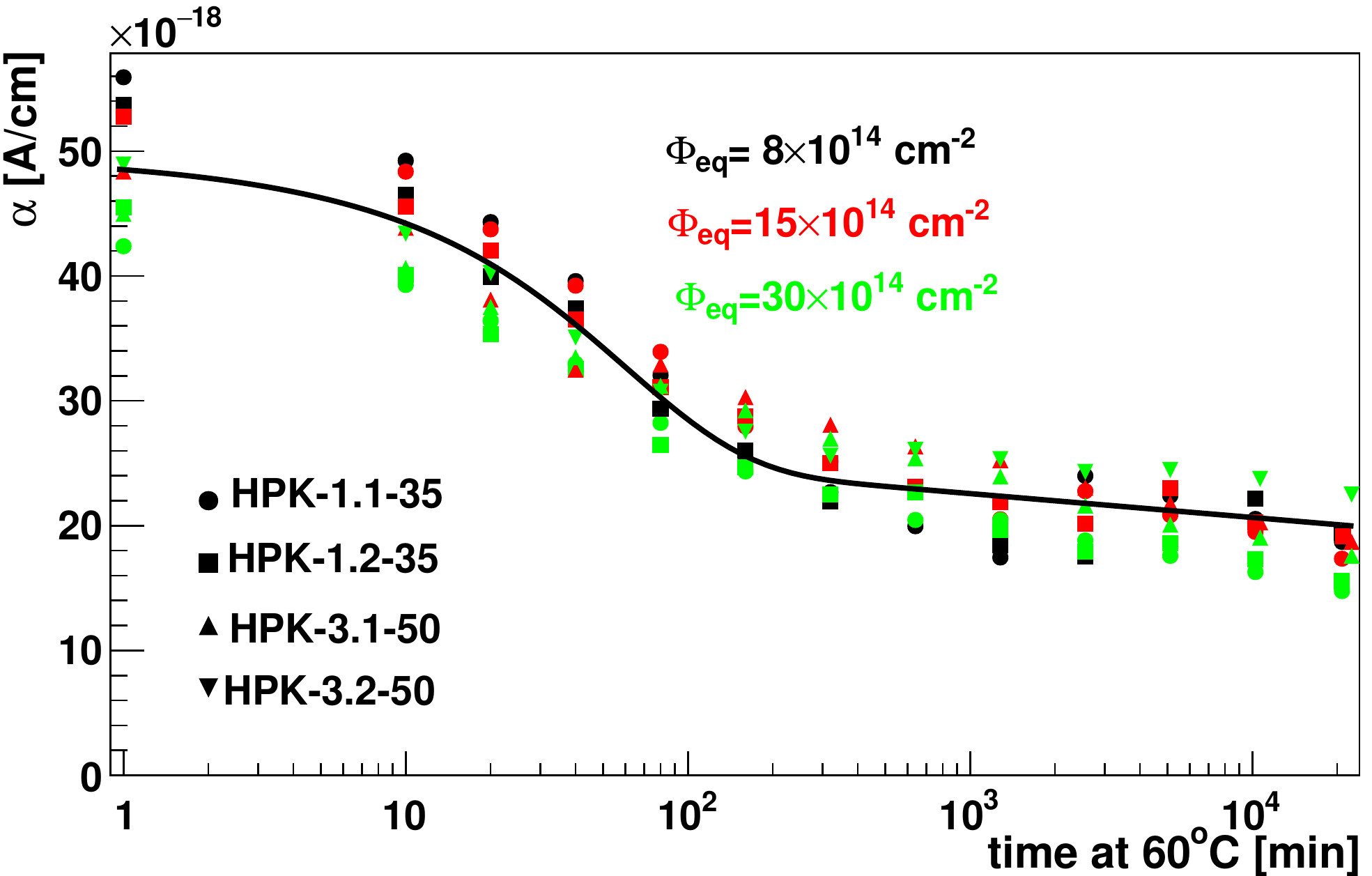,width=0.5\linewidth,clip=} \\
\end{tabular}
\caption{(a) Leakage current measured at $20^\circ$C normalized to sensor volume after 80 min annealing at $60^\circ$C. (b) Dependence of leakage current damage constant on annealing time.
}
\label{fi:IgenFluTime}
\end{figure}

\subsection{Generation current}
The generation current's dependence on fluence for the standard annealing point is shown in
Fig. \ref{fi:IgenFluTime}a. The linear increase with fluence can be observed with the slope $\alpha(20^\circ\mathrm{C},80 \mathrm{min})=3.3\cdot10^{-17}$ cm$^{-1}$, which is lower
than  $\alpha=4 \cdot 10^{-17}$ cm$^{-1}$ commonly used. Although this is a non-negligible difference a combination of several uncertainties, foremost temperature, $I(V_{fd})$ and fluence, can be a reason for it.
The currents normalized to volume and fluence, so called leakage current damage constant $\alpha(20^\circ\mathrm{C}, t)$, are shown in Fig. \ref{fi:IgenFluTime}b. The model used in \cite{Moll97} was fit to the data for all samples, excluding the points for the lowest fluences of 50 $\mu$m devices which showed some gain already at $V_{fd}$
\begin{equation}
\alpha(t)=\alpha_0\,\,\exp(-\frac{t}{\tau_l})+\alpha_1-\alpha_2 \ln(-\frac{t}{\mathrm{1\,\,min}}) \quad .
\label{eq:Ileak}
\end{equation}
The free parameters obtained are gathered in the Table \ref{ta:Igen}.
\begin{table}
\caption{Parameters obtained from the global fit of Eq. \ref{eq:Ileak} to the data shown in Fig. \ref{fi:IgenFluTime}.}
\label{ta:Igen}
\begin{center}
\begin{tabular}{|c|c|c|c|}
\hline
 $\alpha_0 [10^{-17}$ cm$^{-1}$] & $\alpha_1 [10^{-17}$ cm$^{-1}$] & $\alpha_2 [10^{-17}$ cm$^{-1}$] & $\tau_l$ [min]   \\
\hline
 $2.83 \pm 0.22$ & $2.11 \pm 0.27$  & $0.083\pm0.03$  & $60 \pm 13$ \\
\hline
\end{tabular}
\end{center}
\end{table}
The data are in reasonable agreements with previously measured results. The reduction of
generation current with time can be exploited for reduction of the power consumption in operation,
particularly as after completed annealing of the gain layer the charge collection is unaffected by any
further annealing as described in the next section.

\section{Annealing effects on timing and charge collection measurements}

The ultimate benchmark of the annealing influence on the operation of LGADs is its impact on
charge collection and timing resolution. Studies made within ATLAS-HGTD showed that
50 $\mu$m thick devices outperform 35 $\mu$m thick ones, due to larger deposited charge,
smaller capacitance and less steep increase of charge with bias voltage close to the operation point.
Therefore annealing studies in this paper concentrated on HPK-3.1-50 and HPK-3.2-50.

\begin{figure}[!hbt]
\begin{tabular}{cc}
\epsfig{file=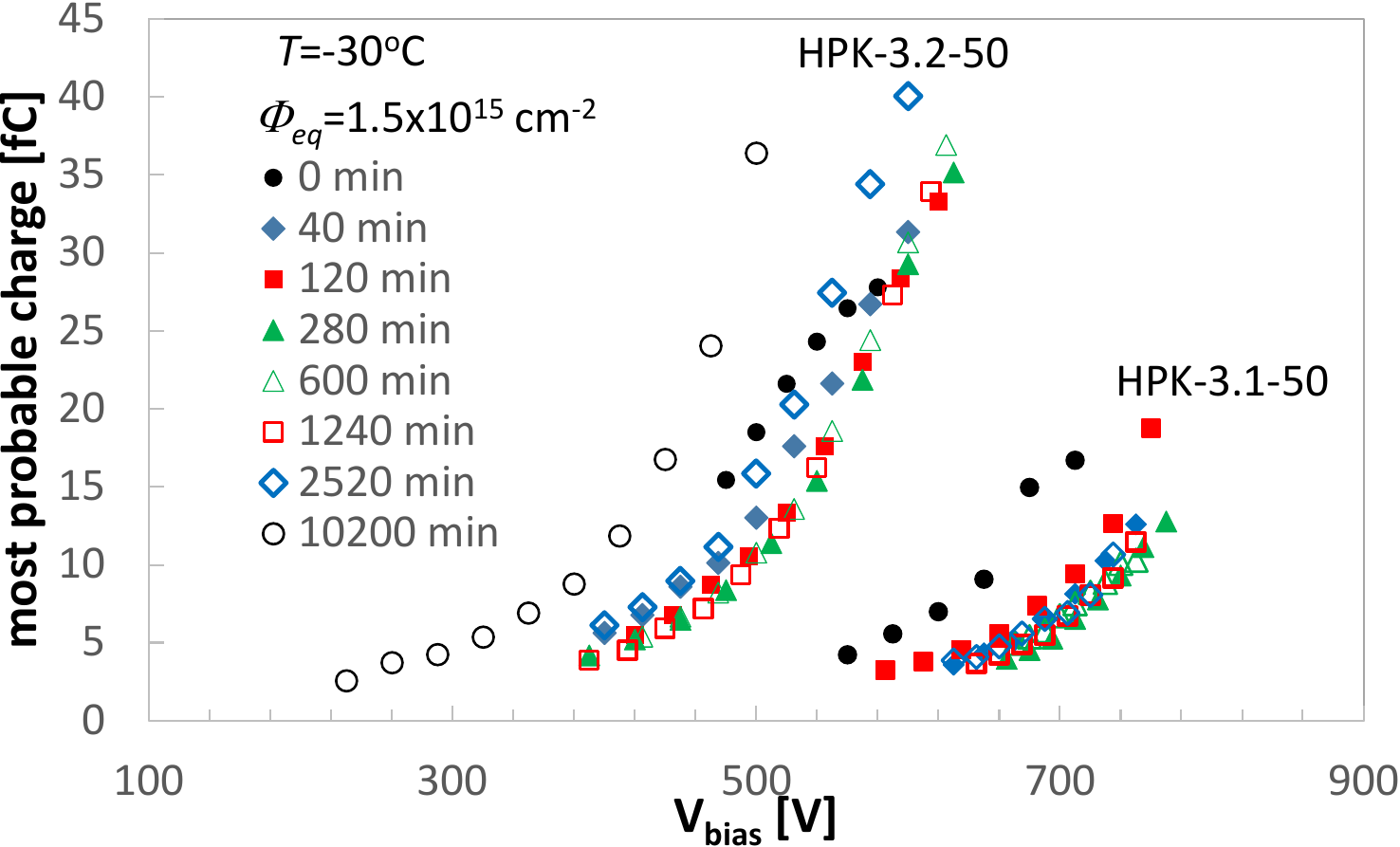,width=0.5\linewidth,clip=} & \epsfig{file=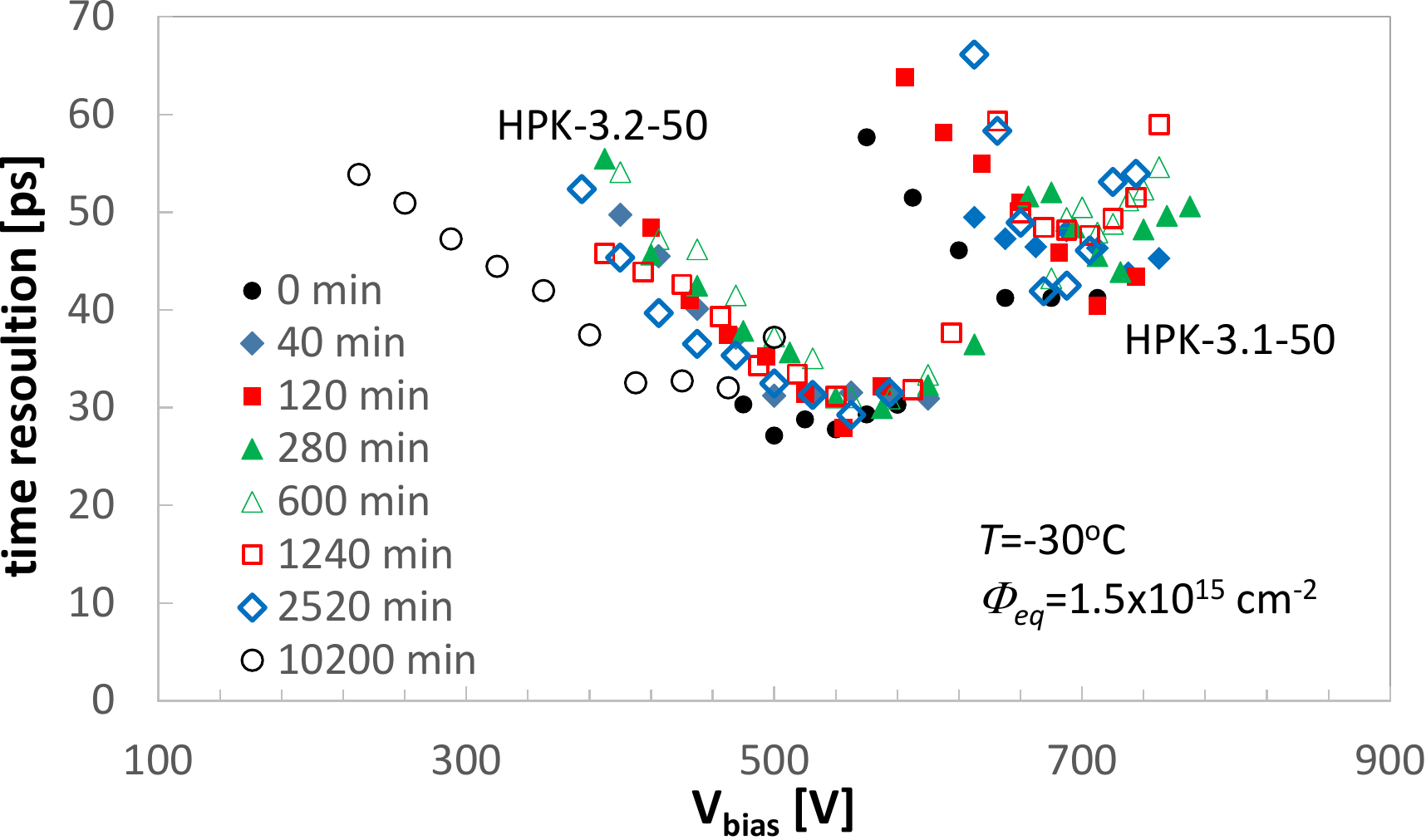,width=0.5\linewidth,clip=} \\
(a) & (b) \\
\end{tabular}
\begin{center}
\epsfig{file=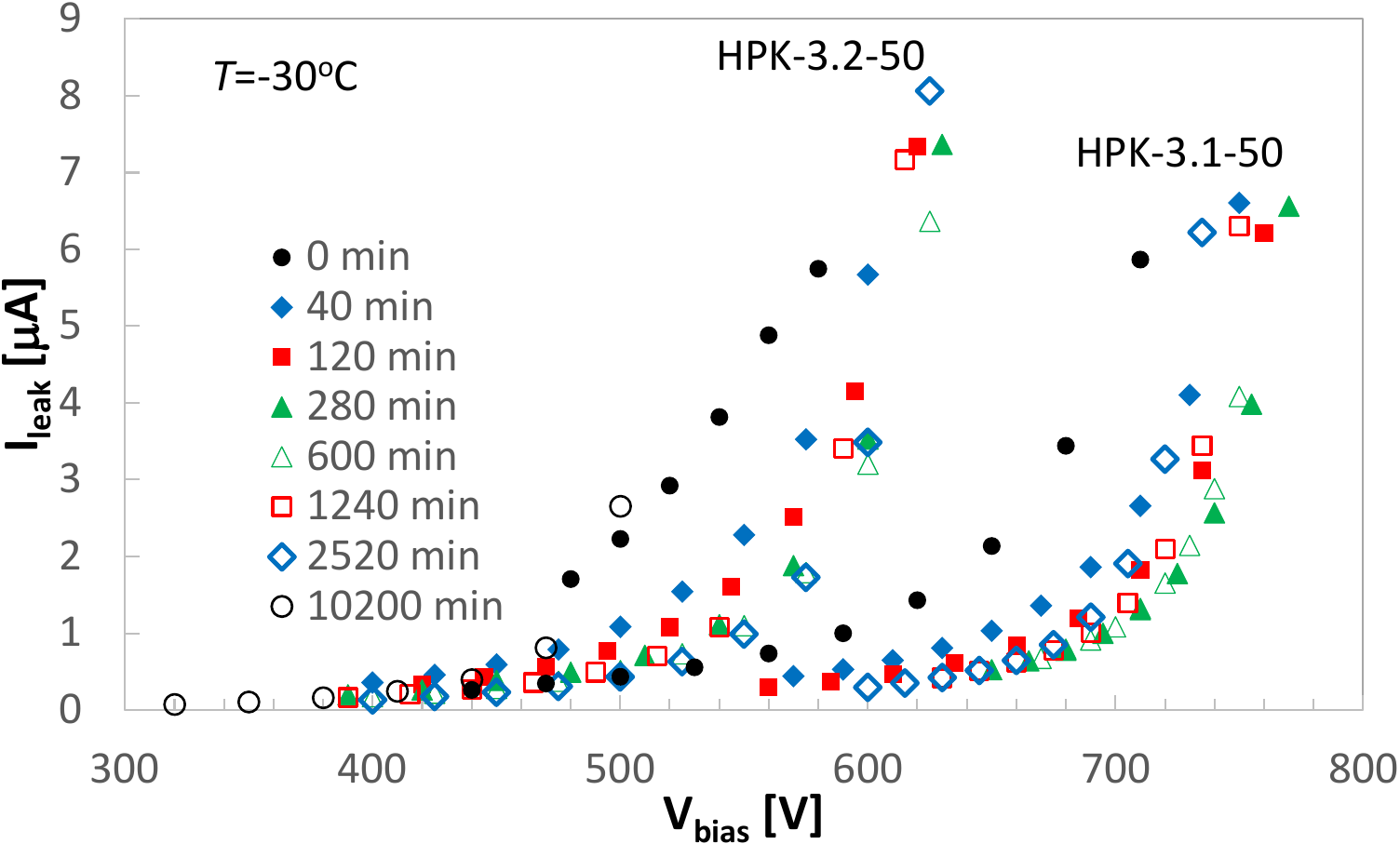,width=0.5\linewidth,clip=} \\
(c) 
\end{center}
\caption{(a) Dependence of most probable charge on bias voltage for 50 $\mu$m thick
detectors irradiated to $\Phi_{eq}=1.5\cdot10^{15}$ cm$^{-2}$ for different annealing times, (b) the corresponding
time resolution and (c) leakage current. The measurements were done at -30$^\circ$C.}
\label{fi:AnnealingT32T31}
\end{figure}

Fig. \ref{fi:AnnealingT32T31}a shows the charge collection (CC) for both sensors
after $\Phi_{eq}=1.5\cdot10^{15}$ cm$^{-2}$. There is an obvious difference in CC which is
due to a different gain layer profile, both in profile depth and dose. The annealing has
little effect, except for the measurement point before any intentional annealing ($t=0$), which
showed significantly larger charge at given voltage than at other annealing points up to 2520 min.
This is in agreement with annealing of $V_{gl}$. Assuming $x_{gl}\approx 1-2$ $\mu$m, the 
difference of 1 V in $V_{gl}$ roughly transfers to 50-25 V difference in operational voltage to get the
same electric field in the gain layer, and hence the same amount of charge. At very large annealing times
of $>10000$ min we noticed a much better performance in CC for HPK-3.2-50 with an indication of increase
already at 2520 min. We had two parallel detectors
and both showed the same behavior. Such a behavior was also observed with ATLAS strip detectors where
substantial increase of charge occurred only after $> 3000$ min annealing \cite{MM2012,LD2019}. This improvement
can be attributed to a larger impact of the bulk on electric field in the gain layer as described in the introduction.

The timing measurements for both detectors are shown in Fig. \ref{fi:AnnealingT32T31}b. The HPK-3.2-50
shows superior performance reaching time resolution of 30 ps. The performance at different annealing times
is in line with the one observed in CC. At the highest voltages the time resolution
deteriorates due to the increase of noise, which leads to the increase of jitter.
The increase of noise was attributed to the measurements with a floating guard ring and the noise disappeared at
later measurements with a grounded guard ring.

The leakage current  measurements for these detectors are shown in Fig.  \ref{fi:AnnealingT32T31}c. Although the sum of bulk and
guard current is shown the contribution of the latter is usually much smaller. The current decreases with annealing
due to decrease of $I_{gen}$ as the gain remains roughly constant except before any intentional annealing and after very long annealing times.

\begin{figure}[!hbt]
\begin{tabular}{cc}
\epsfig{file=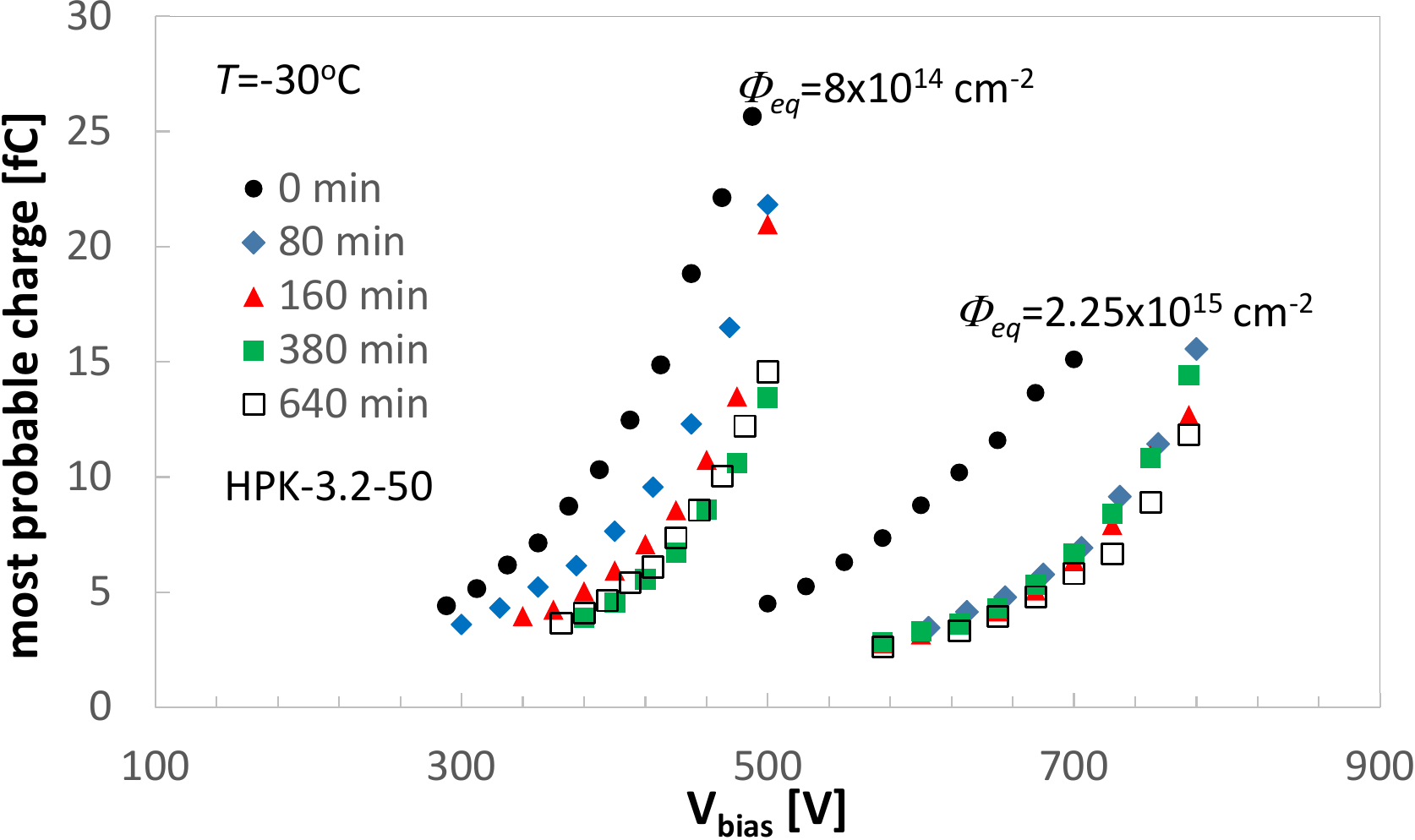,width=0.5\linewidth,clip=} & \epsfig{file=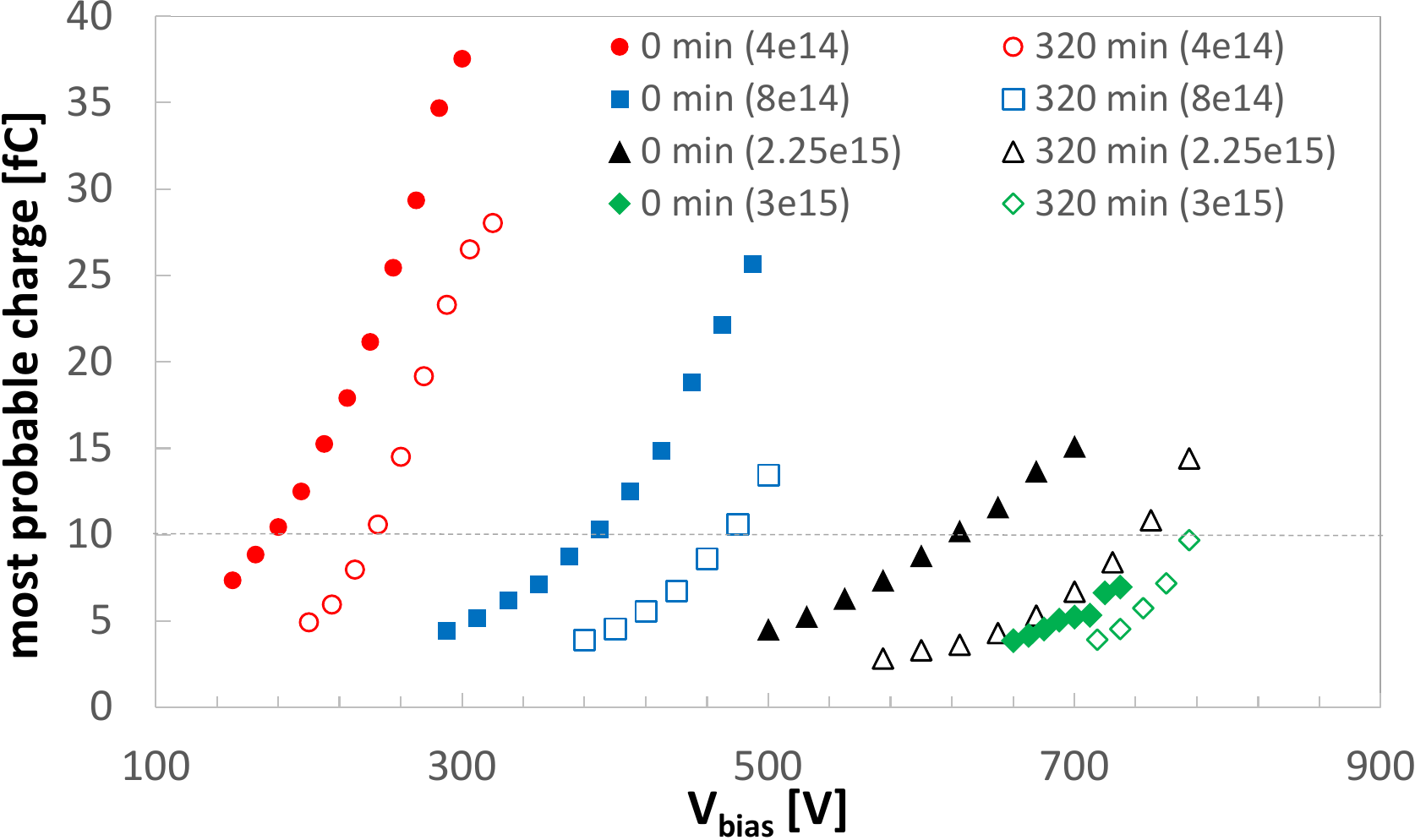,width=0.5\linewidth,clip=} \\
(a) & (b)  
\end{tabular}
\begin{center}
\epsfig{file=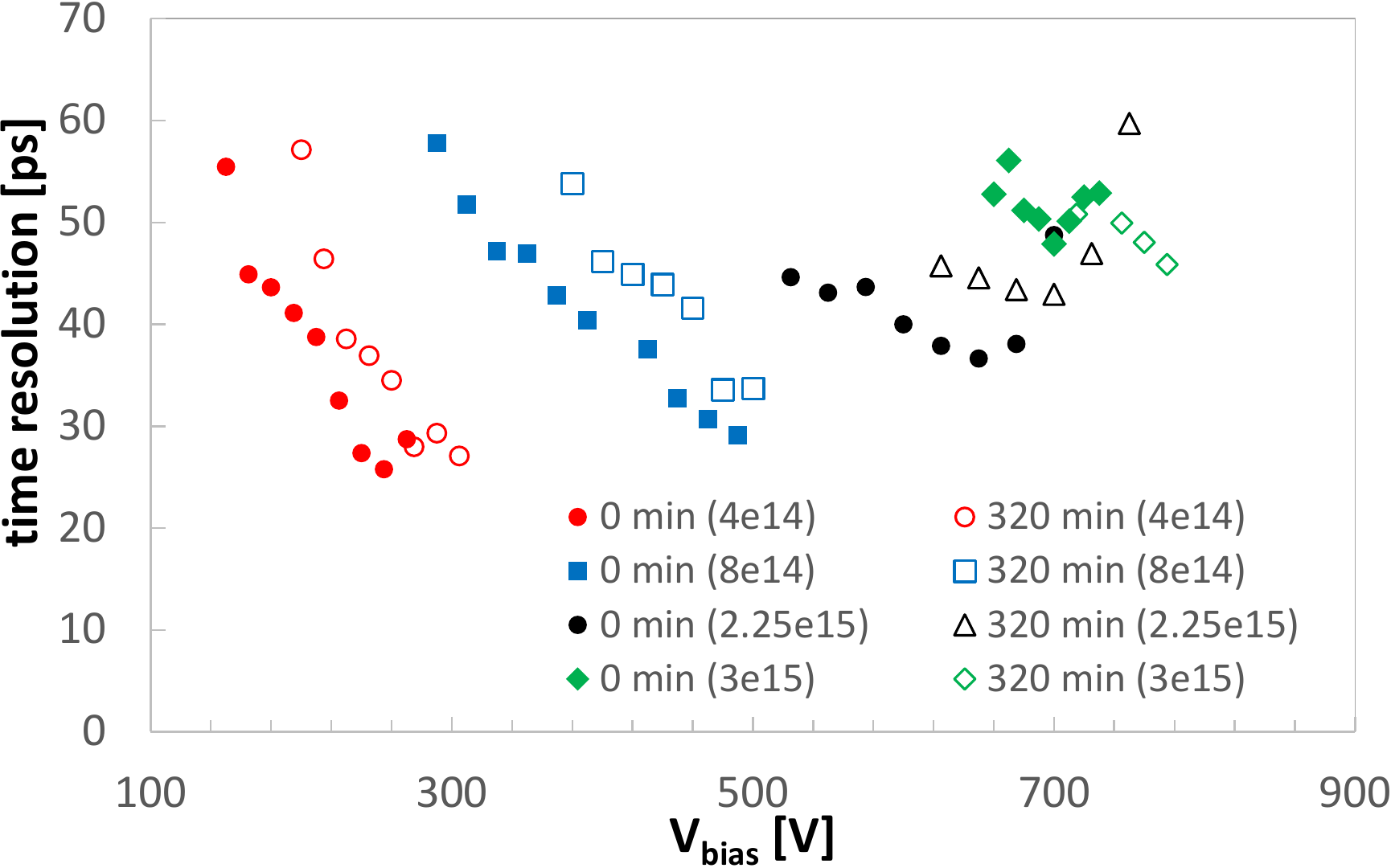,width=0.5\linewidth,clip=} \\
(c) 
\end{center}
\caption{(a) Dependence of most probable charge on bias voltage for HPK-3.2-50
detectors irradiated to two different fluences at different annealing times, (b) The difference
in collected charge between the annealing point of minimum and maximum charge collection for different fluences given in the brackets and (c) the corresponding
time resolution. Measurements were done at -30$^\circ$C.}
\label{fi:OtherAnnPlots}
\end{figure}

The annealing studies at smaller and larger fluences than  $1.5 \cdot 10^{15}$ cm$^{-2}$ show the same behavior
(see Fig. \ref{fi:OtherAnnPlots}a). As most of the studies done so far
were after 80 min annealing, it makes sense to compare the performance of annealed sensors with the one
immediately after irradiation. Such a comparison is shown in Fig. \ref{fi:OtherAnnPlots}b. It is clear
that the bias voltage difference required for collection of e.g. 10 fC (dashed line) between 0 and 320 min annealing time increases slightly
with fluence up to the $\Phi_{eq}=3 \cdot 10^{15}$ cm$^{-2}$, which is in agreement with the increase of $F$ with fluence. At the maximum fluence shown the voltage required
for multiplication is already so high that it comes close to the breakdown, where the bulk multiplication starts to take place and this relation breaks down. It is therefore clear that the standard annealing point for most LGAD studies so far presents a conservative estimate of their timing and charge collection performance.

The time resolution at different fluences is in accordance with charge collection measurements
(see Fig. \ref{fi:OtherAnnPlots}c). The resolution decreases steadily with fluence from 27 ps at $=4\cdot10^{14}$ cm$^{-2}$ to around 50 ps at $3\cdot10^{15}$ cm$^{-2}$.
\subsection{Qualitative explanation of observed results}

\begin{figure}[!hbt]
\begin{center}
\epsfig{file=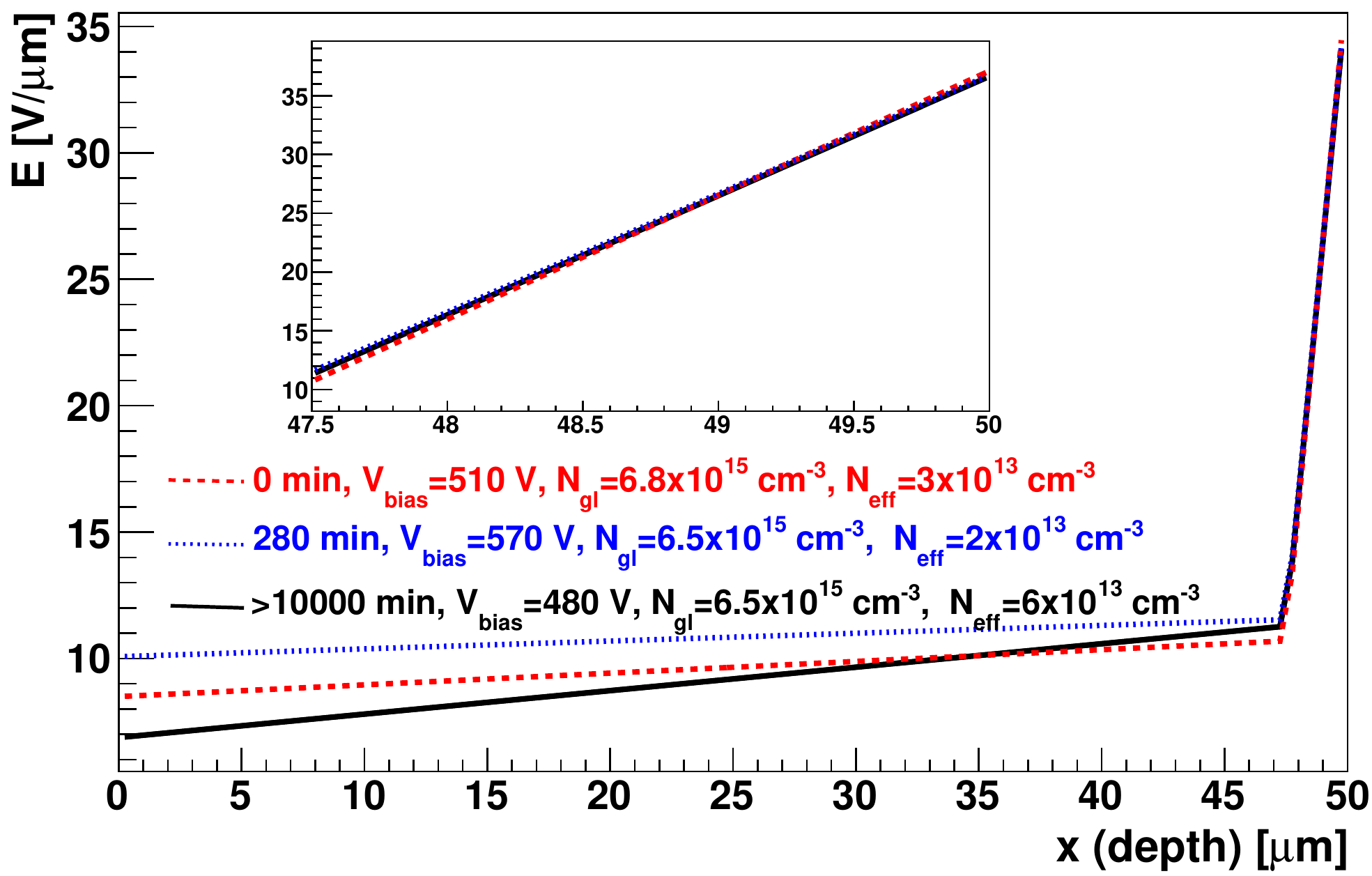,width=0.65\linewidth,clip=}   
\end{center}
\caption{Calculated electric field assuming constant doping concentration and abrupt
transition between the bulk and the gain layer at $x=47.5\,\,\mu$m ($x_{gl}=2.5\,\,\mu$m). The
parameters used for calculation are given in the plot. The inset shows the electric
field in the gain layer.}
\label{fi:EField}
\end{figure}

The effects of initial short term and long term annealing are best illustrated by the calculated electric field
for HPK-3.2-50 detector irradiated to $1.5 \cdot 10^{15}$ cm$^{-2}$. As shown in Fig. \ref{fi:AnnealingT32T31}a around
20 fC is collected at different annealing stages and bias voltages; 480 V after $>10000$ min, 510 V at 0 min
and 570 V after 280 min. The difference originates from a different
electric field in the gain layer due to both $N_{gl}$ and $N_{eff}$.
As the detector is always highly over-depleted with drift velocity close to saturated, the gain
should depend on the electric field in the gain layer only. The electric field calculated
with the approximation of constant
doping concentration in two regions with abrupt transition between them is shown in Fig. \ref{fi:EField}.
The values of $N_{gl}$ ($N_{gl}=2 \varepsilon \varepsilon_0 V_{gl}/\mathrm{e_0}\,x_{gl}^2$) and $N_{eff}$ as measured
at different annealing stages in this work were used. Although the electric field differs significantly in the bulk it is
identical in the gain layer, which validates equal charge measured. This illustrates that although $N_{eff}\ll N_{gl}$ it can significantly
impact the electric field in the gain layer. 


\section{Conclusions}

Both bulk and gain layer properties change with time after irradiation. The gain layer effective dopant
concentration was found to decrease exponentially with time after the irradiation with time constant of
around 50 min at 60$^\circ$C and 650 min at 40$^\circ$C, yielding the estimation of activation energy
of 1.15 eV. Thus, the effective doping concentration in the gain layer won't anneal until the yearly technical stops at HL-LHC.
The relative fraction of the effective doping concentration in the gain layer that
anneals after irradiation increases with fluence from around $\sim 2\%$ after $8 \cdot 10^{14}$ cm$^{-2}$, $\sim 7\%$
after $1.5 \cdot 10^{15}$ cm$^{-2}$ to $\sim 13\%$ after $3\cdot10^{15}$ cm$^{-2}$.

The changes
of the bulk doping concentration can be well described by the Hamburg model. The short term and
long term annealing have introduction rates $g_a \sim 0.01 , g_Y \sim 0.03$ cm$^{-1}$ and
annealing times $\tau_a \sim$ tens min, $\tau_Y \sim 3000$ min similar to those observed before in oxygen rich silicon detectors.
The introduction of stable acceptors was difficult to estimate as the number of fluences was not
enough to accurately model it, but it seems to be compatible with previous measurements. The annealing
of generation current exhibit the same behavior as in standard detectors and follows the
NIEL prediction.

Somewhat larger doping concentration in the gain layer immediately after irradiation,
leads to higher charge collection and better timing resolution at a given voltage, offsetting the operational voltage by
50-100 V in the entire fluence range with respect to that at after standard annealing. However, after few tens of minutes at 60$^\circ$C the
performance remains largely unaffected up to around thousand minutes. At
very large annealing times of $>10000$ min a sizable improvement in charge collection and
timing was observed for sensors irradiated to $\Phi_{eq}=1.5\cdot 10^{15}$ cm$^{-2}$. A simple calculation
of the electric field including all measured bulk and gain layer changes in doping concentration validated the
charge collection measurements during different annealing stages. Although very long times are
probably unpractical for HL-LHC they can be exploited in the case of unplanned events.

The standard annealing
point of 80 min at 60$^\circ$C where most of the studies were done so far therefore represents a
conservative estimate in terms of required operation voltage.


\section*{Acknowledgment}

The authors acknowledge the financial support from the Slovenian Research Agency ( program ARRS P1-0135 and project ARRS J1-1699 ).


\end{document}